\documentclass[12pt,preprint]{article}
\pdfoutput=1 
\usepackage{jheppub} 
\usepackage{gensymb}
\usepackage{graphicx,dcolumn,bm,epsfig,cancel,hyperref}
\usepackage{multirow}
\usepackage{amsmath}
\usepackage{amssymb, times}
\usepackage[utf8]{inputenc}
\newcommand{\comment}[1]{{}}
\numberwithin{equation}{section}
\def\beq{\begin{align}}
\def\eeq{\end{align}}
\newcommand{\bi}{\begin{itemize}}
\newcommand{\ei}{\end{itemize}}
\newcommand{\ben}{\begin{enumerate}}
\newcommand{\een}{\end{enumerate}}
\newcommand{\be}{\begin{equation}}
\newcommand{\ee}{\end{equation}}
\newcommand{\bea}{\begin{eqnarray}}
\newcommand{\eea}{\end{eqnarray}}

\def\vo{\mathcal{V}}
\usepackage{tikz}
\usepackage{subcaption}

\def\vo{\mathcal{V}}

\def\bec{\begin{center}}
\def\eec{\end{center}}
\def\beq{\begin{eqnarray}}
\def\eeq{\end{eqnarray}}

\usepackage{color}
\usepackage{ifthen}
\newboolean{editorial}
\setboolean{editorial}{true}
\newcommand{\editorial}[2]{\ifthenelse{\boolean{editorial}}{\textcolor{red}{[\textsf{\textbf{{#1}}}: }\textcolor{blue}{\textsf{{#2}}}\textcolor{red}{]}}{}}

\usepackage{xcolor}

\author[a]{Robert Wiley Deal,}
\author[b]{Leia Barrowes,}
\author[c,d,e]{John T. Giblin, Jr.,}
\author[f,g]{Kuver Sinha,}
\author[h,i]{Scott Watson,}
\author[b]{Fred C. Adams}

\affiliation[a]{\footnotesize Department of Physics, University of Wisconsin, Madison, WI, 53706, USA}
\affiliation[b]{\footnotesize Department of Physics, University of Michigan, Ann Arbor, MI, 48109, USA }

\affiliation[c]{Department of Physics, Kenyon College, Gambier, Ohio 43022, U.S.A.}
\affiliation[d]{Department of Physics/CERCA/Institute for the Science of Origins, Case Western Reserve University, Cleveland, OH 44106, U.S.A.}
\affiliation[e]{Center for Cosmology and AstroParticle Physics (CCAPP) and Department of Physics, Ohio State University, Columbus, OH 43210, U.S.A.}

\affiliation[f]{\footnotesize Department of Physics and Astronomy, University of Oklahoma, Norman, OK, 73019, USA}

\affiliation[g]{Department of Physics and Astronomy, Rice University, Houston, TX 77005, USA}

\affiliation[h]{\footnotesize Department of Physics, Syracuse University, Syracuse, NY 13244, USA}
\affiliation[i]{\footnotesize Department of Physics and Astronomy, University of South Carolina, Columbia, SC 29208, USA}

\emailAdd{wileydeal@wisc.edu}
\emailAdd{giblinj@kenyon.edu}
\emailAdd{kuver.sinha@ou.edu}
\emailAdd{gswatson@syr.edu}
\emailAdd{barrowes@umich.edu}
\emailAdd{fca@umich.edu}

\title{Cosmological Moduli and Non-perturbative Production of Axions}

\abstract{Cosmological moduli generically come to dominate the energy density of the early universe, and thereby trigger an early matter dominated era. Such non-standard cosmological histories are expected to have profound effects on the evolution and production of axion cold dark matter and dark radiation, as well as their prospects for detection. We consider moduli-axion couplings and investigate the early history of the coupled system,  considering closely the evolution of the homogeneous modulus field, the back-reaction from the axion, and the energy densities of the two fields. A particular point of interest is the enhancement of axion production from modulus decay, due to  tachyonic and parametric resonant instabilities, and the implications of such production on the cosmological moduli problem, axion dark radiation, and the available parameter space for axion dark matter. Using an effective field theory approach, WKB-based semi-analytical analysis, and detailed numerical estimates of the co-evolution of the system, we evaluate the expected decay efficiency of the modulus to axions. The effects of higher-order operators are studied and implications for UV-complete frameworks such as the Large Volume Scenarios in Type IIB string theory are considered in detail.
}

\begin{document}
\maketitle
\flushbottom

\section{Introduction}

The QCD axion and axion-like particles (ALPs) \cite{Weinberg:1977ma,Peccei:1977hh,Wilczek:1977pj,Kim:1979if,Shifman:1979if,Zhitnitsky:1980tq,Dine:1982ah} continue to inspire a vast community effort directed at their discovery as a possible dark matter candidate\footnote{We will refer to both the QCD axion as well as ALPs as ``axions" throughout this work.}. In addition, their cosmological evolution in the early universe is of extreme interest \cite{OHare:2024nmr, Marsh:2015xka}. 
Given the weakness of the interactions between axions and the Standard Model, one can ask in what ways the dark sector may be different and perhaps contain a hint at the underlying UV physics -- particularly given the motivation of the QCD axion as the leading candidate to address the Strong CP problem.
Especially important is the possible connection of the axion to String Cosmology \cite{Cicoli:2023opf} and non-standard cosmological histories inspired by string theory \cite{Batell:2024dsi, Allahverdi:2020bys, Kane:2015jia}. 
Intimately tied to the evolution of the axion are questions of moduli stabilization \cite{McAllister:2023vgy, Allahverdi:2013noa,Allahverdi:2014ppa,Cicoli:2022uqa,Dutta:2009uf,Acharya:2008bk,Acharya:2009zt} and the string landscape \cite{Gendler:2023kjt, Demirtas:2021gsq, Broeckel:2021dpz}.  
Early matter-dominated eras, resulting from string moduli, have been extensively studied in recent years within the context of their effect on axion cold dark matter searches \cite{Batell:2024dsi, Allahverdi:2020bys,Kane:2015jia, Giudice:2000ex, Gelmini:2006pw}, axion dark radiation \cite{Cicoli:2022uqa,Baer:2022fou}, and axion dark matter substructures \cite{WileyDeal:2023trg, Nelson:2018via, Visinelli:2018wza, Blinov:2019jqc}.

A standard way to investigate the  cosmological evolution of the axion in the context of early matter domination is to incorporate the new equation of state within the time evolution of the scale factor, while ignoring the effects of possible axion-moduli couplings \cite{Hertzberg:2008wr, Poulin:2018dzj,Visinelli:2009kt, Baer:2011uz,  Visinelli:2018wza, Hlozek:2017zzf}.  In the standard treatment, the non-standard cosmological epoch appears as a novel temporal dependency of the Hubble constant which also affects the time evolution of the axion.
The axion relic density can also be affected through entropy produced by the decay of the modulus. 
However, if one were to account for the moduli-axion couplings -- in other words -- consider alternative histories caused by a field that is non-trivially coupled to the axion, the situation can have far richer phenomenology \cite{Batell:2024dsi,Giblin:2017wlo,Co:2017mop}. 
Such a coupling affects the equations of motion of both the axion and modulus and, depending on the details of this effect, can result in an altered evolution that impacts both the modulus and the axion. 
Perhaps the most important effects are the possible amelioration of the cosmological moduli problem due to tachyonic and resonant axion production from the modulus -- thus ending the early matter dominated era far sooner than the scale expected from perturbative decay -- and the implications of such production for axion dark matter and dark radiation.

The purpose of this paper is to  study  the early evolution of the axion in non-standard cosmologies with the inclusion of moduli-axion couplings. A particularly simple setting in which to consider such effects is  effective field theory  -- specifically to study the effects of a dimension-five operator. Dimension-five operators coupling the axion to the modulus are motivated by string theory, where axions are a generic expectation from the reduction of higher-dimensional gauge fields \cite{Svrcek:2006yi}.
The non-trivial coupling between the modulus and axion is suppressed by the UV scale of the effective theory\footnote{Depending on the model one considers, this can be set by the Planck or string scale. Following \cite{Cicoli:2016olq}, we restrict ourselves to the Planck scale as a cutoff in this work -- although our results can be applied straightforwardly to lower cutoff scales.}. 
We study in detail both tachyonic and parametric resonance instabilities, which can enhance axion production in a different manner compared to standard perturbative production.
The implications of such production for the Cosmological Moduli Problem are a central focus of our investigation. 
In particular, the question arises: can the enhanced production of axions from the non-perturbative decay of the modulus ``ameliorate" the moduli problem, in the sense of preventing its possible interference with the successes of Big Bang Nucleosynthesis (BBN)?

Given that the background radiation redshifts at a much faster rate than moduli fields, which redshift like cold matter, such instabilities must be extremely efficient in order to reduce the energy density of the modulus to a subdominant component of the cosmology until the scale of its perturbative decay.
We find that tachyonic instabilities that show up at the dimension-five level appear promising to produce such an efficient decay, although once higher-order terms are taken into account these tachyonic instabilities may not persist, depending on the UV origin of the corrections.
Outside of this particular regime in dimension-five treatment, we find that although a large enhancement of axion modes can otherwise occur, it appears highly unlikely to produce a sufficiently efficient modulus decay to affect its later evolution.
However, once one considers a full UV theory, we find that such an efficient decay may be possible if one has a sufficiently strong coupling between the modulus and axion -- far beyond what can be accurately described at the dimension-five level. 

Along the way, we study in detail the cosmological evolution of the coupled modulus-axion system in an expanding universe, first through a semi-analytical treatment employing the WKB approximation, and then through detailed numerical estimates. We carefully track the evolution of the homogeneous modulus field, the magnitude of the back-reaction from the axion field, and the average energy densities of the two fields. We illustrate the enhancement in the axion field and provide a qualitative picture of the decay efficiency of the modulus, and note that our results are sensitive to the initial conditions. A more precise treatment, which includes the full lattice results to calculate the decay efficiency, is reserved for upcoming work. We also draw qualitative conclusions on how our results change once higher-dimensional couplings between the modulus and the axion are taken into account.

It should be emphasized that there is a fundamental distinction between our scenario and models where the axion has couplings to other fields, for example, models of gauge preheating and inflation \cite{Leedom:2024qgr,Adshead:2019lbr,Adshead:2018doq,Adshead:2017xll,Adshead:2015pva}. 
This was also studied in \cite{Adshead:2024ykw}, where it was shown through lattice computations that for an inflaton-axion system, such a resonance could transfer nearly all of the inflaton's energy into the axion field.
Here our focus is slightly different as we are examining the effects of cosmological moduli (and the cosmological moduli problem) on the parametric production of axions -- post-inflation.
A similar case was studied in \cite{Co:2017mop}, where the axion was instead coupled to the radial component of the Peccei-Quinn (PQ) symmetry breaking field.
Although our investigation is motivated by the presence of post-inflationary moduli, we stress that our work may also applicable to scenarios of inflationary reheating provided a suitable (i.e. radiation or matter dominated) background expansion, and so long as the oscillations of the inflaton about its minimum can be well-approximated by retaining only the mass term in the potential.
It should also be noted that the possible amelioration of the cosmological moduli problem has been studied by a subset of the present authors in the context of moduli-gauge field couplings \cite{Giblin:2017wlo}. 
Again, our focus in this paper goes further and investigates not only the effect on the decay of the modulus, but also the evolution of the axion, and, in particular, the coupled modulus-axion system in a post-inflationary setting.

This work is organized as follows. In Section \ref{CMP}, we review the cosmological moduli problem, its motivation, and how it arises in fundamental approaches to early universe cosmology. 
We also qualitatively describe the behavior of axions produced through rapid non-perturbative decay at times close to BBN.
In Section \ref{AMD}, we present the dynamics between the axion and moduli in a simple toy model, and determine whether parametric and tachyonic resonances can both efficiently populate axions while also eliminating the presence of moduli to address the cosmological moduli problem. 
In Section \ref{AUV}, we extend our analysis to general UV completions and provide benchmark cases to study how particular UV completions can affect the results implied by our toy model.
We also demonstrate how such phenomenology is realized in string theory, focusing on the particular case of Large Volume Scenario (LVS) models, which provide a `toolbox' for calculations.
In these models, the K\"ahler modulus perturbatively decays in part to ultra-light closed string axions. 
It can also decay to the QCD axion, realized as the phase of a charged open string field. 
In both cases, the precise coupling of the modulus to the axion can be obtained, and the effect of the modulus on the thermal evolution of the axion can be worked out in detail as we present. 
We also include in the Appendix \ref{appendix} analytical treatment of the dynamics of the modulus and axion zero modes, given the role of the axion zero mode as a motivated cold dark matter candidate.

\begin{table}[]
    \centering
    \caption{\large Guide to Parameters}

    \fbox{     
    \begin{tabular}{l l}
        $R(t)$   
        &
        FLRW scale factor 
    \\
        $\phi$
        &
        Canonically normalized modulus field 
    \\
        $\phi_0$
        &
        Initial modulus zero-mode amplitude 
    \\
        $a$
        &
        Canonically normalized axion field 
    \\
        $a_0$
        &
        Initial axion zero-mode amplitude 
    \\
        $\tilde{k}$ $\equiv \frac{k}{R}$
        &
        Physical momentum 
    \\
        $z\equiv m_\phi t$ 
        &
        Time (normalized to modulus frequency)
    \\
        $\beta \equiv c \phi_0 / m_P$
        &
        Effective gravitational coupling
    \\
        $\mu \equiv m_\phi / m_a$
        &
        Mass ratio 
    \end{tabular}}
    \label{tab:definitions}
\end{table}

\section{Non-perturbative dynamics and the Cosmological Moduli Problem \label{CMP}}

To illustrate the implications of a rapid non-perturbative modulus decay, it is worthwhile to first briefly review\footnote{For more comprehensive reviews of this topic, see e.g. \cite{Allahverdi:2020bys,Kane:2015jia,Cicoli:2023opf}.} the standard picture of the cosmological moduli problem.  During the inflationary period, a light modulus field $\phi$ is displaced far from its minimum, gaining an initial amplitude which is expected to be of the order of the Planck scale \cite{Dine:1995kz,Dine:1995uk,Cicoli:2016olq}.
A minimally-coupled modulus then obeys the equation of motion of a damped harmonic oscillator,
\begin{equation}
    \ddot{\phi}
    +
    3
    H
    \dot{\phi}
    +
    m_\phi^2
    \phi
    =
    0
    .
\end{equation}
While the Hubble scale $H$ is very large (i.e. $H \gtrsim m_\phi$), the modulus field is overdamped and therefore pinned at its initial field value by Hubble friction.
Once the inflaton decays and produces an initial thermal population, it is typically assumed that the universe is radiation-dominated for a brief period.
During this period, the Hubble scale becomes sufficiently low (i.e. $H \lesssim m_\phi$) that the modulus field becomes underdamped and begins to behave similarly to cold matter.
Since cold matter dilutes more slowly than the background radiation, the modulus it will inevitably overtake the energy density of the universe due to its large initial energy, resulting in an extended period of early matter domination \cite{Erickcek:2015jza, Fan:2014zua,Coughlan:1983ci,deCarlos:1993wie}.
This early matter dominated period lasts until the modulus decays (at $H \sim \Gamma_\phi$, where $\Gamma_\phi$ is its perturbative decay width).
Such a decay produces a massive amount of entropy and, in effect, resets the particle content of the universe.
Due to the large entropy production, any pre-existing relics are significantly diluted, and most (if not all) of the modulus decay products will rapidly thermalize and ``reheat'' the thermal background to a temperature set by the scale of the decay.
The decay scale, set by the perturbative decay width $\Gamma_\phi$, is extremely small as the modulus is expected to only acquire gravitational couplings \cite{WileyDeal:2023sry, Bae:2022okh,Cicoli:2012aq,Higaki:2012ar,deCarlos:1993wie}, i.e.
\begin{equation}
    \Gamma_\phi 
    \simeq
    \frac{c}{48\pi}
    \frac{m_\phi^3}{m_P^2}
    \sim
    1.2
    c
    \times
    10^{-27}
    \left(
        \frac{
            m_\phi
        }{
            10\text{ TeV}
        }
    \right)^3
    \text{ GeV}
\end{equation}
where $c \gtrsim \mathcal{O}(1)$ is a model-specific parameter set by the primary decay channels.
The modulus is therefore expected to be long-lived, and could potentially decay during or after Big Bang Nucleosynthesis (BBN) -- erasing the overwhelming successes of BBN \cite{Coughlan:1983ci,Kawasaki:2017bqm,Kawasaki:1999na,Kawasaki:2000en}.
This is the conventional cosmological moduli problem, which is typically evaded by taking the modulus to be sufficiently massive, $m_\phi \gtrsim \mathcal{O}(10-50 \text{ TeV})$, with a precise lower bound being model-dependent.

In models where the modulus decays to a dark matter particle such as a WIMP, the presence of the late-decaying modulus can also drastically alter the produced abundance \cite{Acharya:2008bk,Acharya:2009zt,Allahverdi:2013noa,Allahverdi:2020uax,Bae:2022okh,Cicoli:2022uqa,Baer:2023bbn}.
If the scattering rate of the dark matter particle is large, and the decay temperature of the modulus is below the freeze-out temperature of the dark matter particle, there can be a sizeable enhancement of the dark matter abundance relative to the standard freeze-out mechanism.
This enhancement can bring models with thermally-underproduced dark matter into accord with the measured abundance \cite{Moroi:1999zb,Gelmini:2006pq,Allahverdi:2013noa}.
Conversely, if the scattering rate of the dark matter particle is very small, the dark matter abundance is identical to that produced by the modulus decay (and any relevant cascade decays) -- assuming that the entropy produced by the decay is sufficient to render any pre-existing relics negligible.
This scenario can reproduce the measured abundance in models which have thermally-overproduced dark matter if the net branching ratio is sufficiently low
\cite{Cicoli:2022uqa,Allahverdi:2013noa,WileyDeal:2023trg}.
Additionally, if the modulus does not decay into the dark matter particle (such as if the decay is kinematically forbidden) but the pre-existing abundance is sufficiently large, the entropy produced from the modulus decay can dilute the dark matter abundance to the measured abundance \cite{Allahverdi:2020uax}.

There are also several cosmological connections between moduli and the gravitino.  
As gravitinos are only gravitationally coupled, they have both a highly suppressed interaction rate $\langle \sigma_{3/2} v \rangle$ (see e.g. \cite{Moroi:1995fs}) and an extremely long lifetime (see e.g. \cite{Kohri:2005wn}).
If produced in sufficiently large quantities -- such as during inflationary reheating -- their abundance will persist and, if unstable but sufficiently long-lived, interfere with the successful predictions of BBN when they decay \cite{Kawasaki:2017bqm, Kohri:2005wn,Kawasaki:2004qu,Kawasaki:2008qe,Moroi:1995fs}.
This is the classic gravitino problem.
In the context of a modulus-dominated universe, however, the entropy dilution from modulus decay renders the thermally-produced gravitino population negligible \cite{Bae:2022okh,Kohri:2004qu} and thus eases the constraints on the inflationary sector \cite{Kawasaki:2008qe,Kawasaki:2017bqm}.
On the other hand, if the modulus decay to gravitinos is kinematically accessible, a sizeable gravitino population can be produced by modulus decay which may cause a resurgence of this problem \cite{Hashimoto:1998mu,Kohri:2004qu,Endo:2006zj,Asaka:2006bv,Nakamura:2006uc,Dine:2006ii}. 
This is the modulus-induced gravitino problem.

Additional challenges arise for models with light moduli which may also decay into axions.
As axions are expected to be extremely light, any population of axions which is produced through a perturbative decay of the modulus would be highly relativistic.
Due to the inherent shift symmetry of the axion, any additional interactions are extremely suppressed and thus they are expected to behave as dark radiation (DR) which is highly constrained by CMB measurements.
This has been widely explored in explicit string constructions for the closed string axions of the Large Volume Scenario \cite{Cicoli:2012aq,Higaki:2012ar,Allahverdi:2014ppa,Hebecker:2014gka,Cicoli:2015bpq,Cicoli:2022fzy} and the fibered Large Volume Scenario \cite{Angus:2014bia,Cicoli:2018cgu,Cicoli:2022uqa}, wherein typically the production of dark radiation is sufficiently large to require either a high SUSY-breaking scale (which increases the branching ratio to visible sector particles through the Giudice-Masiero term) or introduce additional field content to the SM so that the current experimental bounds $\Delta N_{\text{eff}} \lesssim 0.29$ \cite{Planck:2018jri} can be satisfied.
This has also been explored in various effective theory models inspired by string constructions \cite{Baer:2023bbn, Baer:2022fou,Higaki:2013lra,Acharya:2015zfk} which similarly find a large -- and potentially problematic -- production of dark radiation unless the branching ratio of the modulus to the visible sector can be made suitably large.

Even if excessive dark radiation production is avoided, the modulus decay can interfere with the axion dark matter abundance produced through the misalignment mechanism.
If the homogeneous component of the axion field begins oscillating before the modulus has decayed perturbatively, the entropy produced by modulus decay can substantially dilute the relic density of misalignment axions \cite{Fox:2004kb,Baer:2023bbn,WileyDeal:2023sry,WileyDeal:2023trg}.
While this dilution may be compensated for with additional tuning of the axion misalignment angle\footnote{This assumes the misalignment angle is a free parameter, which is not the case if the relevant PQ symmetry is broken after inflation.
In this case, the misalignment angle assumes an average value $\langle a_0^2 \rangle \sim \pi^2/3$ \cite{Marsh:2015xka}.}, the tuning required for the misalignment axions to saturate the measured dark matter abundance may be severe -- tentatively counteracting the original theoretical motivation for the axion.
It was also shown in \cite{WileyDeal:2023trg} that the formation of axion miniclusters favors regions of parameter space where the misalignment axion density is highly diluted from entropy production.

However, the standard picture described above can change drastically if the modulus undergoes a rapid, non-perturbative decay into an axion field, analogous to the preheating process in inflationary models.
Based on previous works which study the kinetic preheating process in inflationary models \cite{Adshead:2023nhk,Adshead:2024ykw,Lachapelle:2008sy},
such a decay is expected to be most pronounced within the first few oscillation cycles of the modulus, as the field quickly dilutes during subsequent oscillations due to Hubble expansion.
If the modulus decays completely through this non-perturbative effect, the lower mass bound on $m_\phi$ is effectively erased\footnote{
The lower bound on $m_\phi$ is erased so long as $m_\phi > H_{\text{BBN}} \sim 10^{-24}$ GeV.  
Given that most models lack any symmetry to produce such a light modulus mass, we do not consider this bound any further.
} due the high scale of the decay, which prevents any overlap with BBN.
Additionally, in this case the modulus may not produce a sizable contribution to the entropy of the universe.
Since the modulus would transfer its energy into the axion field, and since any scattering involving the axion is suppressed by the (potentially very large) PQ scale $f_a$ as $\langle \sigma v \rangle_{X \leftrightarrow Y+a} \sim \mathcal{O}(f_a^{-2})$ (see e.g. \cite{Salvio:2013iaa}), the axion field should not thermalize significantly.
Thus, if the modulus decays entirely through the kinetic preheating process, the universe's radiation background is simply set by the inflationary reheating process -- leaving any inflationary relics present, despite the short-lived appearance of a modulus.
The modulus would not ``reheat'' the universe, although it would leave behind an additional axion background.
As we will demonstrate in the next subsection, the axions produced through this decay would translate to a small production of dark radiation or an enhancement of the axion cold dark matter population, depending on the model details.

\subsection{Behavior of axions produced non-perturbatively}

Given the potential for such a drastic shift in the resulting cosmology, it is worthwhile to explore the qualitative behavior of the axions produced through a complete non-perturbative modulus decay.
Specifically, here we are interested in determining if the axions produced will behave similarly to cold dark matter or dark radiation at times close to the BBN epoch.
From this, we can infer how the late-time behavior of axions produced non-perturbatively differs from the dark radiation production and dark matter dilution that may occur within the standard perturbative picture.

Our first task is to estimate the boundary between relativistic and non-relativistic regimes for the axion.
For axions with physical momenta $\tilde{k}^2 \equiv k^2 / R^2$, we expect ``cold'' modes to have $\tilde{k}^2 \lesssim m_a^2$, while ``hot'' modes will obey $\tilde{k}^2 \gtrsim m_a^2$.
Making this comparison at the time of BBN, we can estimate the momentum which divides the hot and cold regimes by
\begin{equation}
    m_a^2
    =
    \tilde{k}_{\text{BBN}}^2
    =
    \frac{k^2}{R_{\text{Decay}}^2}
    \left(
        \frac{R_{\text{Decay}}}{R_{\text{BBN}}}
    \right)^2
    =
    \frac{k^2}{R_{\text{Decay}}^2}
    \left(
        \frac{H_{\text{BBN}}}{H_{\text{Decay}}}
    \right)
\end{equation}
where we have assumed the universe to be radiation-dominated between the modulus decay and BBN.
Given that we expect the modulus to decay non-perturbatively within the first few oscillation cycles, we take $H_{\text{Decay}} \sim m_\phi$.
Adopting the reasonable benchmark values $m_\phi = 10$ TeV and $m_\phi / m_a = 10^{13}$, we estimate that axions produced through the non-perturbative modulus decay will be relativistic and thus behave as dark radiation at the time of BBN if
\begin{equation}
    \frac{k^2}{m_\phi^2}
    \gtrsim
    4.4
    \times
    10^{1}
    \left(
        \frac{
            10^{13}
        }{
            m_\phi / m_a
        }
    \right)^2
    \left(
        \frac{
            m_\phi
        }{
            10 \text{ TeV}
        }
    \right)
    \left(
        \frac{
            2.3 \times 10^{-24} \text{ GeV}
        }{
            H_{\text{BBN}}
        }
    \right)
\end{equation}
while modes produced with smaller $k^2/m_\phi^2$ will be sufficiently redshifted to behave like cold dark matter.

For comparison, in the perturbative decay case we would expect $H_{\text{Decay}} \sim \Gamma_\phi$, which can reasonably be expected to be $\Gamma_\phi \sim 10^{-27}$ for a 10 TeV modulus \cite{WileyDeal:2023sry}.
Axions produced through a perturbative modulus decay are then expected to behave as dark radiation if:
\begin{equation}
    \frac{k^2}{m_\phi^2}
    \gtrsim
    4.4
    \times
    10^{-30}
    \left(
        \frac{
            10^{13}
        }{
            m_\phi / m_a
        }
    \right)^2
    \left(
        \frac{
            \Gamma_\phi
        }{
            10^{-27} \text{ GeV}
        }
    \right)
    \left(
        \frac{
            2.3 \times 10^{-24} \text{ GeV}
        }{
            H_{\text{BBN}}
        }
    \right)
    .
\end{equation}
Given that one would expect $k \sim m_\phi / 2$ in the perturbative decay, the production of dark radiation in the perturbative case is clearly present for phenomenologically expected values.
These estimates provide a key distinction between the two decay processes: while the perturbative case seems to necessarily produce axions which behave as dark radiation, the non-perturbative case may produce axions which can be redshifted to sufficiently low momenta to behave as cold dark matter at BBN -- potentially in place of, or in addition to -- dark radiation, depending crucially on the model's mass hierarchy, $m_\phi / m_a$.
It was also suggested in \cite{Jaeckel:2021gah} that if measured, the dark radiation spectrum (which depends on the production mechanism) could be used as a probe of the reheating process.

Given this distinction, we are now interested in estimating the amount of dark radiation which is produced in the non-perturbative case.
Assuming an instantaneous modulus decay, the energy density of the axions produced from non-perturbative modulus decay is given by:
\begin{equation}
    \rho_a(H_{\text{Decay}})
    \simeq
    \mathcal{B}_{\phi \rightarrow a}
    \,
    \rho_\phi(H_{\text{Decay}})
    \simeq
    \frac{1}{2}
    \mathcal{B}_{\phi \rightarrow a}    
    m_\phi^2
    \phi_0^2
\end{equation}
where we define $\mathcal{B}_{\phi \rightarrow a}\leq 1$ to be the ratio of energy transferred from the modulus to relativistic axions (which in turn depends on the efficiency of the decay, and the percentage of produced axions which are relativistic), and we have assumed the modulus decays within its first oscillation so that it is not significantly diluted due to Hubble expansion.
As we are assuming here that the produced axions are relativistic and therefore dilute as $R^{-4}$, and assuming that there are no sources of entropy production once the modulus has decayed, we can write the axion energy density at a later time $H < H_{\text{Decay}}$ as
\begin{equation}
    \rho_a(H)
    \simeq
    \frac{1}{2}
    \mathcal{B}_{\phi \rightarrow a}    
    m_\phi^2
    \phi_0^2
    \left(
        \frac{
            g_{*S}(T) T^3
        }{
            g_{*S}(T_{\text{Decay}}) 
            T_{\text{Decay}}^3
        }
    \right)^{4/3}
    .
\end{equation}
We are now in a position to estimate the amount of dark radiation produced through a non-perturbative modulus decay, characterized by the effective number of neutrinos $\Delta N_{\text{eff}} = \rho_a / \rho_\nu$, where $\rho_\nu = \frac{7}{8} \frac{\pi^2}{15} \left(\frac{4}{11}\right)^{\frac{4}{3}} T^4$ is the energy density of a single neutrino species (see e.g. \cite{Conlon:2013isa,Kolb:1990vq}).
Under the assumption that the non-perturbative decay occurs at the oscillation temperature of the modulus, we have \cite{Bae:2022okh,WileyDeal:2023sry} $T_{\text{Decay}}\sim \sqrt{m_P m_\phi}$ so that we can estimate $\Delta N_{\text{eff}}$ as\footnote{
It is worth noting that the conventional method used to estimate $\Delta N_{\text{eff}}$ in the perturbative case \cite{Cicoli:2012aq,Conlon:2013isa,Higaki:2012ar,Cicoli:2022uqa,Acharya:2015zfk} is not valid here, since it assumes $\rho_{\text{SM}}$ (and thus $\rho_\nu$) is populated by only $\phi$ decay, while in the non-perturbative case $\rho_{\text{SM}}$ is left over from inflationary reheating.
}
\begin{equation}
    \Delta N_{\text{eff}} 
    \simeq
    \frac{60}{7\pi^2}
    \left(
        \frac{11}{4}
    \right)^{4/3}
    \mathcal{B}_{\phi \rightarrow a}    
    \left(
        \frac{
            \phi_0
        }{
            m_P
        }
    \right)^2
    \left(
        \frac{
            g_{*S}(T)
        }{
            g_{*S}(T_{\text{Decay}}) 
        }
    \right)^{4/3}
    .
\end{equation}
Taking the conservative estimates for the entropic degrees of freedom $g_{*S}(T_{\text{BBN}}) \lesssim 4$ and $g_{*S}(T_{\text{Decay}}) \gtrsim 100$, we see that
\begin{equation}
    \Delta N_{\text{eff}} 
    \simeq
    0.046
    \,
    \mathcal{B}_{\phi \rightarrow a}    
    \left(
        \frac{
            \phi_0
        }{
            m_P
        }
    \right)^2
    \left(
        \frac{
            g_{*S}(T_{\text{BBN}})
        }{
            4
        }
    \right)^{4/3}
    \left(
        \frac{
            100
        }{
            g_{*S}(T_{\text{Decay}}) 
        }
    \right)^{4/3}
    \ll 
    0.29
\end{equation}
and as we expect $\mathcal{B}_{\phi \rightarrow a} \leq 1$ and $\phi_0 / m_P \lesssim 1$, we find that the amount of dark radiation produced through the non-perturbative decay is easily within the current bounds $\Delta N_{\text{eff}} \lesssim 0.29$ from Planck 2018 \cite{Planck:2018jri}.
Indeed, the estimated upper bound on the produced $\Delta N_{\text{eff}}$ is still below the sensitivity of the proposed CMB-S4 experiment which is projected to detect an effective number of neutrinos as low as $\Delta N_{\text{eff}} \sim 0.06$ \cite{Abazajian:2019eic}.

From this estimate, we find a stark contrast between the cosmologies predicted in the perturbative versus the non-perturbative modulus decays.
In the non-perturbative case, dark radiation production is naturally suppressed independent of the details of couplings to any additional sectors.
This is primarily due to the scale at which the decay occurs in this scenario -- since the modulus decays before it overtakes the energy density of the background radiation present from inflationary reheating, the produced dark radiation will only constitute a small correction to the overall radiation density and thus a small $\Delta N_{\text{eff}}$.
In the perturbative case however, the original inflationary radiation bath has redshifted significantly by the time of modulus decay.
The background radiation is then sourced (or reheated) by the modulus decay, which makes the ratio of visible radiation to dark radiation directly produced from modulus decay the determining factor for $\Delta N_{\text{eff}}$.
Additionally, in the non-perturbative case some (or all) of the axions produced through modulus decay may behave as cold dark matter, thus reducing the expected dark radiation production further while also enhancing the dark matter abundance.

\subsection{Partial non-perturbative decays}

Given the large separation of scales between the onset of modulus oscillations and its perturbative decay, one would imagine a highly efficient resonance is required so that any energy density remaining in the modulus after the preheating process does not still overtake radiation -- albeit at a much later time when compared to the fully perturbative case.
Despite the expectation of a high efficiency in this process \cite{Adshead:2023nhk,Adshead:2024ykw}, our aim here is to determine the efficiency which divides the dominantly perturbative and dominantly non-perturbative regimes in a more quantitative sense -- especially given the substantial differences in the cosmological predictions between the two.

Defining an efficiency factor to be $\kappa_{\phi \rightarrow a} \leq 1$, we can approximate the energy left in the modulus immediately after the non-perturbative decay as
\begin{equation}
    \rho_\phi 
    (H \sim m_\phi)
    \simeq
    \frac{1}{2}
    \left(
        1
        -
        \kappa_{\phi \rightarrow a}
    \right)
    m_\phi^2 
    \phi_0^2
    .
\end{equation}
As we are now interested only in the case where the radiation and modulus energy densities are comparable at the perturbative decay scale, we assume that the universe is radiation-dominated between $m_\phi \gtrsim H \gtrsim \Gamma_\phi$.
The modulus energy density at the scale of the perturbative decay can then be estimated as
\begin{equation}
    \rho_\phi 
    (H \sim \Gamma_\phi)
    \simeq
    \frac{1}{2}
    \left(
        1
        -
        \kappa_{\phi \rightarrow a}
    \right)
    m_\phi^2 
    \phi_0^2
    \left(
        \frac{
            \Gamma_\phi
        }{
            m_\phi
        }
    \right)^{3/2}
    .
\end{equation}
The radiation energy density at the perturbative decay scale can then be written in terms of the perturbative decay width $\Gamma_\phi$ as
\begin{equation}
    \rho_r 
    (H \sim \Gamma_\phi)
    \simeq
    3
    m_P^2
    \Gamma_\phi^2
    .
\end{equation}
With these two estimates for the radiation and modulus energy densities, we find that the two are comparable ($\rho_r \sim \rho_\phi$) for an efficiency given by
\begin{equation}
    \kappa_{\phi \rightarrow a}
    \sim 
    1
    -
    6
    \frac{
        m_P^2    
    }{
        \phi_0^2    
    }
    \left(
        \frac{
            \Gamma_\phi        
        }{
            m_\phi        
        } 
    \right)^{1/2}
    .
\end{equation}
Using our benchmark values, we see that the efficiency which divides the dominantly perturbative and dominantly non-perturbative regimes is given roughly by
\begin{equation}
    \kappa_{\phi \rightarrow a}
    \sim 
    1
    -
    1.9
    \times 
    10^{-15}
    \left(
        \frac{
            m_P^2    
        }{
            \phi_0^2    
        }
    \right)
    \left(
        \frac{
            \Gamma_\phi        
        }{
            10^{-27} \text{ GeV}
        }
    \right)^{1/2}
    \left(
        \frac{
            10\text{ TeV}
        }{
            m_\phi        
        }
    \right)^{1/2}
    .
\end{equation}
It is worth noting here that since the decay width is expected to scale as $\Gamma_\phi \propto m_\phi^3$, the above efficiency scales as $\kappa_{\phi \rightarrow a} -1 \propto -m_\phi$.
Heavier moduli masses therefore allow for less efficient decays, however we then expect $m_\phi \gtrsim \mathcal{O}(10^{14} \text{ TeV})$ in order for a decay with an efficiency $\kappa_{\phi \rightarrow a} \lesssim 0.9$ to be dominantly non-perturbative.
The key takeaway here is that a near-perfect efficiency appears to be required in order to circumvent the cosmological moduli problem via the non-perturbative decay process.

\section{Axion-Moduli Dynamics \label{AMD}}
With sufficient phenomenological motivation in hand, we now investigate the dynamics of the modulus coupled to an axion.
We work with a flat, FLRW background metric\footnote{We work in Planck units so that $m_P^2 = 8\pi/G$.}
\be{ds^2=dt^2 - R(t)^2 d\vec{x}^2}
\ee
where we use $R(t)$ to denote the scale factor.
To begin, we present the toy model action which adequately encapsulates the gravitational coupling of interest:
\begin{equation}
    \label{eq:action}
    S
    =
    \int 
    d^4x 
    \,
    \sqrt{-g}
    \, 
    \left(
        \frac{1}{2}
        \partial_\mu \phi 
        \partial^\mu \phi 
        +
        \frac{1}{2}
        \partial_\mu a
        \partial^\mu a 
        +        
        \frac{c}{m_P}
        \,
        \phi 
        \partial_\mu
        a
        \partial^\mu 
        a
        -
        \frac{1}{2}
        m_\phi^2
        \phi^2
        -
        \frac{1}{2}
        m_a^2
        a^2
    \right)
    .
\end{equation}
A few comments are in order.
First, in an effective field theory one would expect a tower of higher-dimensional terms coupling $\phi$ to the axion kinetic term.
Second, this toy action can be derived in numerous string models, such as both the open and closed string axions in a Type IIB setting \cite{Cicoli:2022uqa,Cicoli:2012aq}, 
as well as in the $G_2$-MSSM in an $M$-theory setting \cite{Acharya:2015zfk}.
Third, we note that although the characteristic axionic shift symmetry $a \rightarrow a + \text{constant}$ is violated by the mass term in this form, we will discuss the implications of additional potential terms which restore this symmetry in Section~\ref{AUV} -- and thus why we are justified in neglecting any additional couplings here up to dimension-five.

Throughout this work, we find it useful to use a time variable normalized to the modulus oscillation frequency,
\begin{equation}
    z \equiv m_\varphi t.    
\end{equation}
For reference, we summarize all relevant parameters in Table~\ref{tab:definitions}.
The covariant Euler-Lagrange equations of motion from \eqref{eq:action} are given by
\begin{equation}
    \label{eq:eomModulus}
    \phi^{\prime \prime}
    -
    \frac{1}{m_\phi^2 R^2}
    \nabla^2
    \phi 
    +
    \frac{
        3
        H    
    }{
        m_\phi
    }
    \phi^\prime
    +
    \phi
    =
    \frac{c}{m_P}
    \left[ 
        (a^\prime)^2
        -
        \frac{1}{m_\phi^2 R^2}
        \nabla
        a
        \cdot
        \nabla
        a
    \right]
\end{equation}
and
\begin{equation}
    \label{eq:eomAxion}
    a^{\prime \prime}
    -
    \frac{1}{m_\phi^2 R^2}
    \nabla^2
    a
    +
    \frac{
        3
        H    
    }{
        m_\phi
    }
    a^\prime
    +
    \frac{m_a^2/m_\phi^2}{1+2\frac{c}{m_P}\phi}
    a
    =
    -
    2
    \frac{c/m_P}{1+2\frac{c}{m_P}\phi}
    \left[ 
        \phi^\prime
        a^\prime
         -
        \frac{1}{m_\phi^2 R^2}
        \nabla
        \phi 
        \cdot 
        \nabla
        a
    \right]
\end{equation}
where primes denote differentiation with respect to $z \equiv m_\phi t$, and we use $\nabla$ to refer to the spatial gradient.
As expected, these equations reduce to the standard damped harmonic form in the absence of the coupling, $c \rightarrow 0$. 
The Hubble parameter may also be evaluated explicitly as 
\begin{equation}
    \frac{3H}{m_\phi}
    =
    \begin{cases}
        2/z
        &
        \quad
        \text{(Matter dominated)}
        \\
        3/(2z)
        &
        \quad
        \text{(Radiation dominated).}
    \end{cases}
\end{equation}

\subsection{Mode Decomposition}

Having now presented the full system of equations in \eqref{eq:eomModulus}-\eqref{eq:eomAxion}, it is now worth further investigating the qualitative behavior induced by the presence of the gravitational coupling.
To begin, we assume here that the modulus is largely unaffected by the presence of the back-reaction term (as would certainly be the case for $a_0 / \phi_0 \sim \mathcal{O}(f_a / m_P) \ll 1$) and is thus predominantly a homogeneous field.
This assumption holds under our approximations\footnote{Although the effective modulus mass can be driven tachyonic through the kinetic coupling term, we find that this only leads to efficient production of moduli quanta once the back-reaction term becomes large - at which point the approximations in our numeric treatment of the homogeneous modulus component have also broken down.}, where the modulus potential consists only of a (constant) mass term. This is in contrast to works such as e.g. \cite{Barnaby:2009wr, Krippendorf:2018tei}, where additional corrections to the modulus potential can produce significant amounts of moduli quanta even without sizable axion production. Although we expect such corrections could lead to qualitative differences in our results, we leave systematic study of such additional terms to the modulus potential for future work.
Upon mode-expanding the axion field,
\begin{equation}
    \label{eq:modeExpansionDefn}
    a
    =
    \int 
    \frac{
        d^3
        k
    }{
        (2\pi)^3
    }
    \, 
    a_k 
    \exp(-ikx),
\end{equation}
we can rewrite \eqref{eq:eomAxion} as the equation of motion for axions with physical momenta $\tilde{k}^2 \equiv k^2 / R^2$:
\begin{equation}
    \label{eq:axionKModeEOM}
    a_k^{\prime \prime}
    +
    \left(
        \frac{3H}{m_\phi}
        +
        \frac{
            2
            \frac{c}{m_P}
            \phi^\prime
        }{
            1
            +
            2
            \frac{c}{m_P} \phi
        }
    \right)
    a_k^{\prime} 
    +
    \left( 
        \frac{\tilde{k}^2}{m_\phi^2}
        +
        \frac{
            m_a^2/m_\phi^2
        }{
            1
            +
            2
            \frac{c}{m_P}
            \phi
        }
    \right)
    a_k 
    =
    0
\end{equation}
where, again, we have neglected any spatial dependence of the modulus.

Before we proceed, we first note the presence of an additional friction term for the axion that is induced by the modulus.
Before the modulus has begun to oscillate (i.e. $H \gtrsim m_\phi$), we have $\phi' \sim 0$ which renders this contribution negligible during this stage.
Once the modulus has begun to oscillate (i.e. $H\lesssim m_\phi$), this additional term can be $\mathcal{O}(1)$ and -- depending on the sign of $\frac{c}{m_P} \phi^{\prime}$ -- results in either an enhancement or a diminution of the friction for the axion field.
However, since the modulus dilutes as $\phi \propto z^{-1}$ during matter domination and $ \phi \propto z^{-3/4}$ during radiation domination, we expect that the friction term induced by the gravitational coupling quickly becomes subdominant in magnitude to the usual Hubble friction, which dilutes as $H \propto z^{-1}$. 
Thus, if $\frac{c}{m_P} \phi^\prime$ is ``small'' then we expect this term to become inconsequential for the resulting phenomenology, however if the new friction term is ``large'' then a significant enhancement of the axion field can occur before this effect becomes subdominant to Hubble friction.
We will study this in detail in Sections~\ref{sec:mathieuInstabilities}-\ref{sec:numericalEstimates}.

The second additional contribution in \eqref{eq:axionKModeEOM} is a modification\footnote{Were we to include the inhomogeneous component of the modulus here, it would also arise as an additional contribution to the axion mass term. Inspection of \eqref{eq:eomAxion} shows this contribution also has a factor of $(1+2\frac{c}{m_P} \phi)^{-1}$, which also produces a naive tachyonic instability in the axion field.}
to the axion mass term.
The presence of the gravitational coupling can induce a tachyonic instability in the axion field, specifically if $\frac{c}{m_P} \phi < - 1/2$.
If this condition is not met, however, this correction is then expected to be negligible as super-horizon modes will not enter the Hubble radius and begin to oscillate until the modulus field amplitude has been substantially diluted, while sub-horizon modes have $\tilde{k}^2\gg m_a^2$.
As we will see in Section~\ref{AUV}, the appearance of this particular tachyonic instability is a potential indication of pushing our toy model too far, depending on crucial details of the UV completion which may preserve or remediate this instability.

Having briefly explored the qualitative modifications to the axion dynamics, we now return to a more quantitative investigation.
Making the field redefinition 
\be
    \mathit{a}_k\!\left(z\right)
    =
    \mathit{A_k}\!\left(z\right)
    R\! \left(z\right)^{-\frac{3}{2}} 
    \exp
    \left(
        -
        \int_1^z
        \frac{\frac{c}{m_P} \phi^\prime(\xi)}{1+2 \frac{c}{m_P}\phi(\xi)}
        d\xi
    \right),
\ee
\eqref{eq:axionKModeEOM} can be rewritten as the equation of a harmonic oscillator,
\be 
\label{eomaxion}
A^{\prime \prime}_k(z)+\omega_k^2(z)A_k=0,
\ee 
with the complete {\bf time-dependent} frequency given by
\begin{multline} 
\label{eq:wkbFrequency}
\omega^2_k 
= 
-
\Bigg[
    \frac{3}{2}
    \frac{H^\prime}{m_\phi}
    +
    \frac{9}{4}
    \frac{
        H^2
    }{
        m_\phi^2
    }
    + 
    \frac{
        3
        H
    }{
        m_\phi
    }
    \frac{
        \frac{c}{m_P} \, 
        \phi^\prime
    }{
        1
        +
        2\frac{c}{m_P} \phi 
    }
    \\
    + 
    \frac{  
        \frac{c}{m_P} \, 
        \phi^{\prime \prime}
    }{
        1
        +
        2\frac{c}{m_P} \phi 
    } 
    -
    \left(
        \frac{
                \frac{c}{m_P}
                \phi^\prime
        }{
            1
            +
            2\frac{c}{m_P} \phi 
        } 
    \right)^{2} 
    -
    \left(
        \frac{
            \tilde{k}^2             
        }{  
            m_\phi^2
        }
        +
        \frac{
            m_a^2 / m_\phi^2
        }{
            \left(
                1
                +
                2\frac{c}{m_P} \phi 
            \right)            
        }
    \right) 
\Bigg]. 
\end{multline}

This is a complicated equation that involves both gravity and the dynamics of the modulus and the axion. However, as explicit from this equation, dimensional scales matter.
When examining particle production in such a time-dependent background -- like that provided by the scalar and gravitational interactions here -- the WKB approximation has proven a useful analytic tool to estimate particle production (see e.g. \cite{Mukhanov:2005sc,Lawrence_1996,Gubser:2003vk,Kofman:2004yc,Watson:2004aq,Cremonini:2006sx,Greene:2007sa}). 
Given an equation of the form \eqref{eomaxion}, particle production begins when the leading adiabatic invariants are violated \cite{Mukhanov:2005sc}.
Assuming an initial vacuum solution of the form
\begin{equation}
 {A}_k \approx \frac{1}{\sqrt{2 \omega_k(z)}} e^{-i \int^z \omega_{k}(z') dz'},    
\end{equation}
the adiabatic constraints hold as long as
\begin{center}
\be
\frac{{\omega^\prime}}{{\omega}^2} \ll 1 \;\;\;\;\;\;\;\; \frac{{\omega}^{\prime \prime}}{{\omega}}\ll1.
\label{WKBconditions}
\ee
\end{center}
When these conditions are violated, particle production
begins.
Given our field decomposition and the WKB approach to study fluctuations of the axion vacuum state, we can approximate production by considering the field modes in a semi-classical manner.

\subsection{Instabilities and Axion Growth Rates}
\label{sec:mathieuInstabilities}

We now investigate growth rates of the axion due to the presence of the additional friction term, which now appears in our time-dependent frequency \eqref{eq:wkbFrequency}.
Once $H < m_\phi$, the modulus quickly begins oscillating at a rate faster than the cosmological expansion time.
For timescales shorter than a Hubble time, we can parameterize the modulus field as
\be
    \phi
    \simeq
    \phi_0
    \cos z,
\ee
where again $z\equiv m_\phi t$.
Taking the ``late-time'' limit so that $H/m_\phi \ll 1$ and $H^\prime$ terms can be neglected, and assuming $c\phi_0/m_P <1$ so that we may retain only terms up to first-order, we can rewrite \eqref{eq:wkbFrequency} as
\begin{equation} 
\label{eq:wkbFrequencyLateTimes}
\omega^2_k
\simeq
\Bigg[
    \frac{c}{m_P} \, 
    \phi_0
    \cos z
    +
    \frac{
        \tilde{k}^2             
    }{  
        m_\phi^2
    }
\Bigg]
.
\end{equation}
The adiabatic constraints are then violated once 
\begin{equation}
    \frac{
        -
        \frac{
            c
        }{
            m_P
        }
        \phi_0
        \sin z
    }{
        2
        \left[
            \frac{c}{m_P} \, 
            \phi_0
            \cos z
            +
            \frac{
                \tilde{k}^2             
            }{  
                m_\phi^2
            }
        \right]^{3/2}    
    }
    \gtrsim 
    1
    \qquad 
    \frac{
    -
        \frac{
            c
        }{
            m_P
        }
        \phi_0
        \left(
            2 
            \frac{
                \tilde{k}^2             
            }{  
                m_\phi^2
            }
            \cos z 
            + 
            \frac{
                c
            }{
                m_P
            }
            \phi_0
            \left(
                \cos^2 z 
                + 
                1
            \right)
        \right)
    }{
        4 
        \left[
            \frac{
                c
            }{
                m_P
            }
            \phi_0
            \cos z
            +  
                \frac{
                    \tilde{k}^2             
                }{  
                    m_\phi^2
                }
        \right]^{2}
    }   
    \gtrsim
    1
    .
\end{equation}
Under the approximations contained in our frequency \eqref{eq:wkbFrequencyLateTimes}, the comparison of \eqref{eomaxion} to the Mathieu equation is also immediate:
\begin{equation}
    \label{eq:mathieu}
    \frac{
        d^2u
    }{
        dx^2
    }
    +
    \left[
        \mathcal{A}_k
        +
        2q
        \,
        \cos(2x)
    \right]
    u
    =
    0
\end{equation}
with the identifications
\begin{equation}
    x
    \equiv 
    \frac{z}{2},
    \qquad
    \mathcal{A}_k
    \equiv
    4
    \frac{
        \tilde{k}^2
    }{
        m_\phi^2
    },
    \qquad
    q
    \equiv 
    2
    \frac{c}{m_P}
    \phi_0
    .
\end{equation}

\begin{figure}[htb!]
    \centering
    \includegraphics[width=0.49\linewidth]{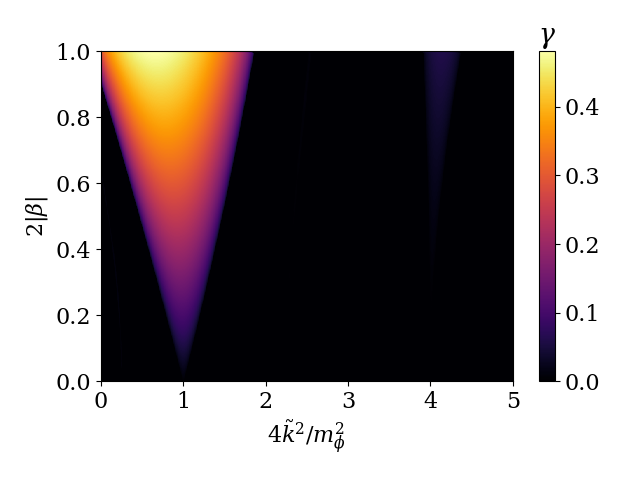}
    \includegraphics[width=0.49\linewidth]{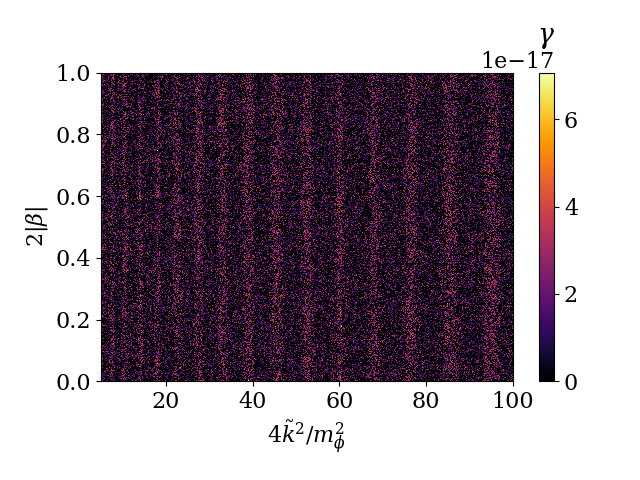}
    \caption{
        Growth rate, $\gamma$, of \eqref{eq:mathieu} for a $q$-domain corresponding to $|\beta|\leq1/2$ and separated $\mathcal{A}_k$-domains, where we have made the definition $\beta \equiv \frac{c}{m_P} \phi_0$. For $\mathcal{A}_k <5$ (left), we see one large band of instability, and $\mathcal{A}_k > 5$ (right) is a stable region with tiny fluctuations from $\gamma=0$ in a distinct pattern.
    }
    \label{fig:mathieu-late-times}
\end{figure}

\begin{figure}[htb!]
    \centering
    \includegraphics[scale=0.5]{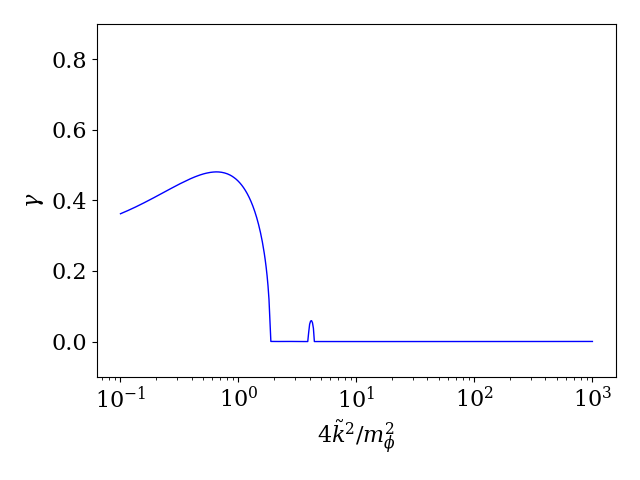}
    \caption{
    Growth rate for the particular choice of $2|\beta|=1$, where we have retained the convenient definition $\beta \equiv \frac{c}{m_P}\phi_0$.}
    \label{fig:mathieu-powerSpectrum}
\end{figure}

Through the identification with the Mathieu equation, we expect some level of instability for various axion modes -- depending on $\mathcal{A}_k$ and $q$.
We show the axion growth rate $\gamma$, which we define as Lyapunov exponent of the equation of motion for $A_k$ (i.e. $A_k \propto e^{\gamma z}$), in Figure~\ref{fig:mathieu-late-times} for a range of $2\beta \equiv 2\frac{c}{m_P} \phi_0 \leq 1$ as a function of the physical momentum, $4\tilde{k}^2/m_\phi^2$.
We see that for modes with $0 \lesssim 4 \tilde{k}^2/m_\phi^2 \lesssim 2$, axion modes can grow with a rate as large as $a_k \propto e^{0.5z}$ for the larger range of $\frac{c}{m_P}\phi_0$ considered.
However, growth is effectively non-existent for modes with $4 \tilde{k}^2/m_\phi^2 \gg 1$ where the growth rate is of the order $\gamma \sim 10^{-17}$.
Figure~\ref{fig:mathieu-powerSpectrum} shows the growth rate of the axion over several magnitudes of momenta, where the second instability band is now noticeable.

It should be stressed that although this simplified treatment provides insight into the axion growth rate at a given time, it does not incorporate its evolution as the universe expands in a straightforward manner.
In particular, as axions with high momenta ($\tilde{k}\gg m_\phi$) redshift, they eventually fall into the largest instability band.
However, by this time the modulus has diluted and the effective $\phi_0$ value is substantially reduced -- again suppressing growth for these modes.
Although these effects are difficult to study analytically, we will see these effects numerically in Section~\ref{sec:numericalEstimates}.

Having investigated the growth rates from the additional friction term present in \eqref{eq:wkbFrequency}, we now shift to investigation of the naive tachyonic instability.
In particular, we note that a tachyonic instability is pronounced for $\frac{c}{m_P}\phi \lesssim -1/2$ while either $\phi^\prime \sim \phi^{\prime \prime} \sim 0$ or $\phi^\prime < 0$.
In the former case, \eqref{eq:wkbFrequency} simplifies to 
\begin{equation}
\label{eq:wkbFrequencyEarlyTimes}
\omega^2_k
\simeq
\Bigg[
    -
    \frac{3}{2}
    \frac{H^\prime}{m_\phi}
    -
    \frac{9}{4}
    \frac{
        H^2
    }{
        m_\phi^2
    }
    +
    \frac{
        \tilde{k}^2             
    }{  
        m_\phi^2
    }
    +
    \frac{
        m_a^2 / m_\phi^2
    }{
        \left(
            1
            +
            2\frac{c}{m_P} \phi_0 
        \right)            
    }
\Bigg].
\end{equation}
Although the second adiabatic condition is extremely lengthy and not particularly illuminating, the first is sufficiently simple to see the relevant effect:
\begin{equation}
\frac{
    -
    \left[
        \frac{3}{4}
        \frac{H^{\prime \prime}}{m_\phi}
        +
        \frac{9}{4}
        \frac{
            H
            H^\prime
        }{
            m_\phi^2
        }
        +
        \frac{H}{m_\phi}
        \frac{
            \tilde{k}^2             
        }{  
            m_\phi^2
        }
    \right]
}{
\left[
    -
    \frac{3}{2}
    \frac{H^\prime}{m_\phi}
    -
    \frac{9}{4}
    \frac{
        H^2
    }{
        m_\phi^2
    }
    +
    \frac{
        \tilde{k}^2             
    }{  
        m_\phi^2
    }
    +
    \frac{
        m_a^2 / m_\phi^2
    }{
        \left(
            1
            +
            2\frac{c}{m_P} \phi_0
        \right)            
    }
\right]^{3/2}
}
\gtrsim
1
.
\end{equation}
As the universe expands and the Hubble friction becomes comparable to the modulus mass, the modulus initial amplitude begins to decrease -- and if $\frac{c}{m_P} \phi_0 \lesssim -1/2$ initially, this contribution begins to blow up.
This can also be seen from the original axion equation of motion, \eqref{eq:axionKModeEOM}, which simplifies to
\begin{align}
    \frac{m_\phi^2}{m_a^2}
    a_k''
    +
    \frac{3H}{m_a}
    \frac{m_\phi}{m_a}
    a_k' 
    + 
    \left(
        \frac{\tilde k^2}{m_a^2} 
        +
        \frac{
            1
        }{
            1
            +
            2
            \frac{c}{m_P}
            \phi
        }
    \right) 
    a_k
    =
    0
    \label{eq:axion-early-time}
\end{align}
before the modulus has begun to oscillate, and where we have normalized time with respect to the frequency of the homogeneous axion mode.

\begin{figure}[htb!]
    \centering
    \includegraphics[width=0.7\linewidth]{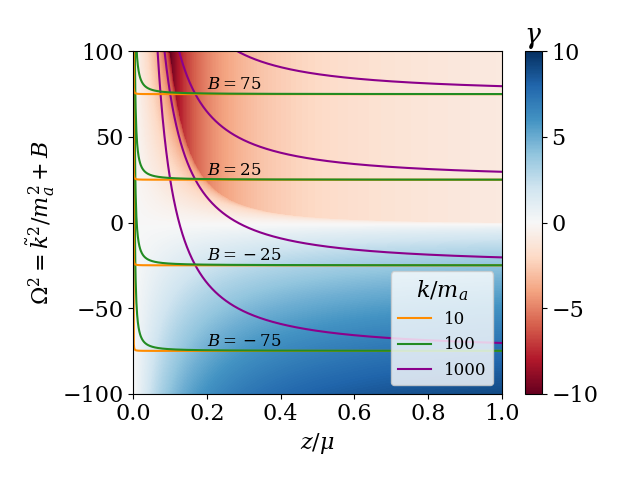}
    \caption{ Growth rate, $\gamma$, as a function of time, normalized to the frequency of the homogeneous axion mode, and an effective frequency depending on the physical wavenumber, $\tilde k$, and $B\equiv (1+2\frac{c}{m_P}\phi)^{-1}$. 
    Here, we have defined $\mu \equiv m_\phi/m_a$.
    The lines are trajectories for fixed comoving wavenumber of the axion, $k=R\tilde k$. 
    The resulting approximate axion solution $a_k\propto e^{\gamma(z)z}$ is exponential with varying growth rate along each trajectory.}
    \label{fig:early-time-stability}
\end{figure}

Figure~\ref{fig:early-time-stability} shows how the axion growth rate, $\gamma$, evolves over time for various effective frequencies $\Omega^2 \equiv \frac{\tilde{k}^2}{m_a^2} + \frac{1}{1+2 \frac{c}{m_P}\phi}$, where $\Omega^2$ also evolves with time.
Using this stability map and some sample trajectories of different co-moving modes, we see that the frequency is eventually dominated by 
$
\Omega^2 \sim (1+2\frac{c}{m_P}\phi)^{-1}
$ as the wavelength redshifts.
However, even for 
$
\frac{c}{m_P} \phi
<-\frac{1}{2}
$, higher-energy modes are less likely to experience exponential growth, and their final abundance will depend on exactly how long this regime lasts.
Of course, this apparent instability may be an artifact of neglecting higher-order terms depending on details of the UV theory, as we will see in Section~\ref{AUV}.

\subsection{Numerical Estimates}
\label{sec:numericalEstimates}
In this section, we present numerical estimates for the evolution of the axion and moduli fields in an expanding universe.
Although these estimates do not provide us with quantitative decay efficiencies, they do provide us with a sense of scale and direct us to the regime of parameter space in which our approximations break down, indicating where a more detailed treatment such as lattice computations are required.

\begin{figure}[htb!]
    \centering
    \includegraphics[scale=0.6]{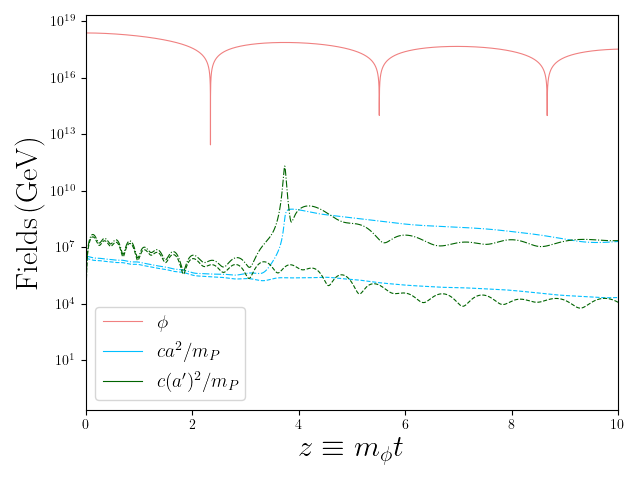}
    \caption{Magnitudes of the modulus field (red) and the back-reaction due to the axion field (blue) and axion velocity (green).
    Dashed curves have $c \phi_0 / m_P = 1.2$ and dot-dashed curves have $c \phi_0 / m_P = 1.6$. Here, we take $a_0/\phi_0 = 10^{-1}$.}
    \label{fig:dim5FieldsLargeXi}
\end{figure}

For the modulus, we solve \eqref{eq:eomModulus} under the assumption that the back-reaction is negligible, and that the modulus is dominated by its homogeneous component, i.e. $\phi = \phi(t)$.
We also adopt the standard initial conditions $\phi_0 = m_P$ and $\phi^\prime_0 = 0$.
For the axion component, we solve \eqref{eq:axionKModeEOM} for several values of $k/m_\phi$ including the homogeneous mode, $k/m_\phi=0$.
From this, we can approximate the value of the axion field through our mode expansion definition, \eqref{eq:modeExpansionDefn}, so that the average variance of the axion field can be written as 
\begin{equation}
    \langle 
        a^2
    \rangle
    =
    \int
    \frac{d^3k}{(2\pi)^3 R^3}
    a_k^2
\end{equation}
with an analogous expression for $\langle (a')^2 \rangle$.
The energy density of the axion can then be straightforwardly computed from the stress tensor, which after averaging over all space produces the expression
\begin{equation}
    \langle
    \rho_a(x)
    \rangle
    \simeq
    \frac{1}{2}
    \int 
    \frac{dk \, k^2}{2\pi^2 R^3}
    \left[
        \left( 
            1
            +
            2
            \frac{c}{m_P}
            \phi
        \right)
        m_\phi^2
        (
        a_k^\prime
        )^2
        +
        \left( 
            1
            +
            2
            \frac{c}{m_P}
            \phi
        \right)
        k^2
        a_k^2
        +
        m_a^2
        a_k^2
    \right]
    .
\end{equation}
For the axion initial conditions, we assume a scale-invariant power spectrum of a Gaussian random field\footnote{The generation of isocurvature perturbations is also of some concern, but as discussed in \cite{Iliesiu:2013rqa} this is model dependent and depends on the cosmological history.} so that 
\begin{equation}
    \langle
    a_k
    \,
    a_{k'}
    \rangle 
    =
    \frac{\pi^2}{2}
    \left(
        \frac{\Delta_s^2 a_0^2}{k^3}
    \right)
    \delta(k-k')
\end{equation}
where $\Delta_s^2 \sim 10^{-10}$
\cite{Planck:2018jri,Giblin:2017wlo}. 
Although these initial conditions suffice for our current purpose of illustrating the enhancement in the axion field and providing a qualitative picture of the decay efficiency, our results are sensitive to the initial conditions.
More precise treatment, which include full lattice results to calculate the decay efficiency, will be the subject of upcoming work \cite{upcomingWork}.

In Figure~\ref{fig:dim5FieldsLargeXi}, we display the evolution of the homogeneous modulus field $\phi$ compared to the magnitude of the back-reactions from the axion field (blue curves) and the axion velocity (green curves) for $a_0 /\phi_0 = 10^{-1}$, corresponding to $f_a \sim 10^{17}$ GeV.
It is immediately apparent that neither contribution to the back-reaction grows to become significant for the values of $c\phi_0/m_P \in \{1.2, \, 1.6\}$ considered here.
We furthermore note that for smaller PQ scales, $f_a \ll 10^{17}$ GeV, we expect even further suppression of the back-reaction under our assumptions for the axion initial conditions, which are proportional to $a_0 \sim f_a$.
Additionally, larger values of $c \phi_0 / m_P $ induce a tachyonic instability where $1+2\frac{c}{m_P}\phi \sim 0$ when the modulus finishes its first oscillation -- 
although this is likely a symptom of pushing our toy model too far, as we will see from our treatment in Section~\ref{AUV}.

\begin{figure}[htb!]
    \centering
    \includegraphics[scale=0.47]    {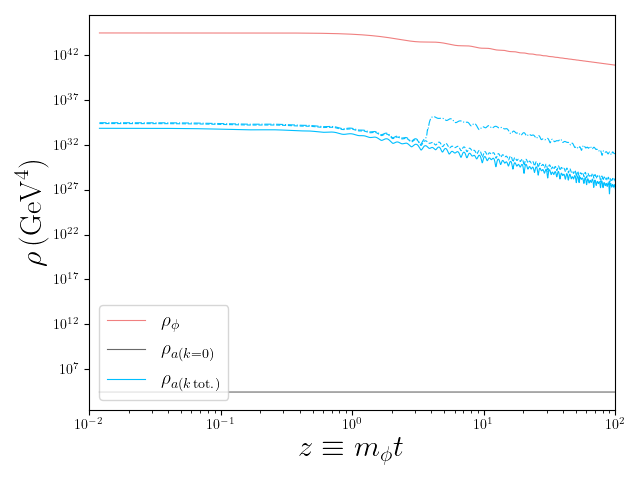}
    \includegraphics[scale=0.47]{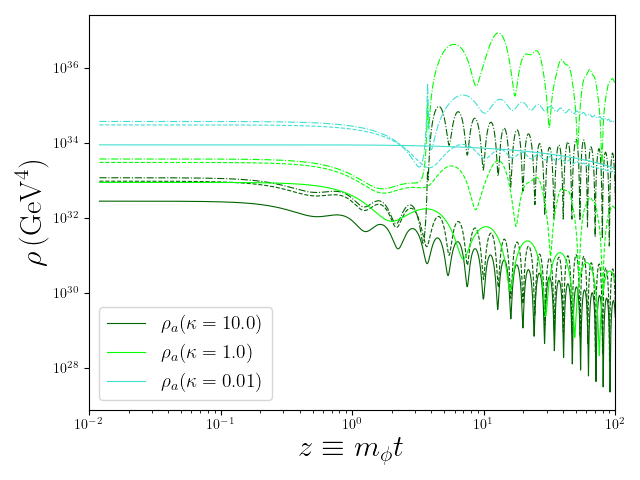}
    \caption{Left: Energy densities of the modulus (red) and axion (blue) fields. We also display the axion zero mode (gray) for comparison. Right: Contributions to the axion energy density from individual modes.
    We define $\kappa \equiv 4\tilde{k}^2 / m_\phi^2$, where $\tilde{k}$ is taken to be the physical momentum at the onset of modulus oscillations. 
    In both plots, solid curves denote the absence of gravitational coupling, while dashed curves have $c\phi_0 / m_P = 1.2$ and dot-dashed curves have $c\phi_0/m_P = 1.6$.
    Both plots take $a_0 /\phi_0 = 10^{-1}$.}
    \label{fig:dim5EnergyLargeXi}
\end{figure}

In Figure~\ref{fig:dim5EnergyLargeXi}, we display the average energy density of the axion field (blue) and the modulus (red) in the left panel. 
We also include the contribution from the lone axion zero-mode (gray) for comparison, where the frozen mode contains only the energy density $\rho_{a_0} \sim m_a^2 a_0^2$ whereas the frozen inhomogeneous modes re-enter the Hubble radius with an energy density $\rho_{a_k} \sim (k^2+m_a^2)a_k^2$, and $k^2 \gg m_a^2$ despite $a_k \sim 10^{-5} a_0$.
In the right panel, we display the contributions of individual modes which have $\kappa \equiv 4\tilde{k}^2 / m_\phi^2 = 0.01$ (blue), $1$ (light green), and $10$ (dark green), where $\tilde{k}$ is taken to be the physical momentum at the onset of modulus oscillations. 
From this panel, we see the absence of the gravitational coupling ($c\phi_0/m_P = 0$, shown by the solid curves) reproduces usual expected behavior of these modes -- once modes re-enter the Hubble radius ($\tilde{k}/H \sim 1$), they begin to oscillate and diminish.
Inclusion of the gravitational coupling then produces amplification of modes, particularly with $4\tilde{k}^2 / m_\phi^2 \sim 1$.
Modes with $4\tilde{k}^2 / m_\phi^2 \sim 0.01$ then see a slightly smaller growth rate when they re-enter the horizon, while modes with $4\tilde{k}^2 / m_\phi^2 \sim 10$ must redshift before receiving any significant growth.
Thus, within the regime where we would expect the dimension-five term to be dominant, we see that although we may achieve some growth of axion modes, it appears hardly sufficient to achieve the near-perfect modulus decay required to eliminate the cosmological moduli problem.
We now turn to study aspects of UV completions, and investigate if these apparent conclusions remain true once higher-order corrections are taken into account -- especially in regimes where higher-order terms contribute substantially.

\section{Axions and UV Completions \label{AUV}}
In the previous section, we exclusively focused on the effects from a dimension-five coupling.
However, in a particular UV completion one would generically expect a tower of higher-dimensional operators also contributing to the equations of motion, which can alter the predictions of the effective field theory outside of the EFT's regime of validity.
In this section, we attempt to parameterize a general UV completion and draw qualitative conclusions on how our results change once higher-dimensional terms are taken into account.
To accomplish this, we begin with the assumption that the axion still respects a global discrete shift symmetry, $a \rightarrow a + 2\pi N$ for $N\in \mathbb{Z}$, which arises through e.g. instanton effects.
For the QCD axion, this assumption is well-motivated.
Although gravitational corrections are not expected to respect global symmetries, a viable Peccei-Quinn solution to the Strong CP problem requires any gravitationally-induced terms present in the action to be suppressed by a factor of at least $m_P^{-8}$ (the well-known `axion quality problem' \cite{Kamionkowski:1992mf,Dine:1992vx}).
On the other hand, there is no a priori reason that gravitational corrections to axions from string theory must respect such a discrete shift symmetry -- so long as they do not play the role of the QCD axion.
Here, we restrict our focus to the case where the discrete shift symmetry is left intact and leave study of higher-order terms which break this symmetry for future work.
We do not consider corrections to the modulus potential, which would appear only as anharmonic terms in the modulus equation of motion.

From these considerations, we can write a general action with the inclusion of higher-order corrections, and study the implications for general UV completions.
The first set of corrections we include parametrizes the tower of gravitational interactions induced from the UV.
The second set of corrections we consider includes a general axion potential which respects a discrete shift symmetry.
The action which includes both sets of corrections can thus be written as 
\begin{equation}
    \label{eq:actionHigherOrderGravitational}
    S
    \simeq
    \int 
    d^4x 
    \,
    \sqrt{-g}
    \, 
    \left(
        \frac{1}{2}
        \partial_\mu \phi 
        \partial^\mu \phi 
        -
        \frac{1}{2}
        m_\phi^2
        \phi^2
        + 
        \sum 
        \limits_{n=0}^\infty
        \left(
            \frac{\phi}{m_P}
        \right)^n
        \left[
            c_n
            \partial_\mu
            a
            \partial^\mu 
            a
            +
            d_n
            \cos 
            \left(
                a
                /
                f_a
            \right)
        \right]
    \right)
\end{equation}
where $c_n$ and $d_n$ are the respective $n$-th order couplings to the axion kinetic and potential terms, and may determined from a given UV completion.
Regardless of the UV completion, the requirement of a canonically normalized axion fixes $c_0 = 1/2$.
It is worth noting that although here the $c_n$ are dimensionless, the $d_n$ have mass dimension-four and are predominantly set by the cutoff scale for the non-perturbative effects which provide the axion its periodic potential (see e.g. \cite{Marsh:2015xka} and references within).
The axion mass term is then obtained via Taylor expansion, $m_a^2 \equiv d_0 / f_a^2$.
From this identification, we would then naively expect that we have, in general, $d_n \sim \mathcal{O}(m_a^2 f_a^2)$.

The equations of motion corresponding to the above action are given by
\begin{equation}
    \label{eq:modulusEOMUV}
    \phi^{\prime \prime}
    -
    \frac{1}{m_\phi^2 R^2}
    \nabla^2
    \phi 
    +
    \frac{
        3H
    }{
        m_\phi
    }
    \phi^{\prime}
    +
    \phi
    = 
    \frac{
        f_a^2
        \,
        h(\phi)
    }{m_P}
    \cos(a/f_a)
    +
    \frac{
    g(\phi)
    }{m_P}
    \left[
        (a^{\prime})^2
        -
        \frac{1}{m_\phi^2 R^2}
        \left(
            \nabla
            a
        \right)^2
    \right]
\end{equation}
and
\begin{multline}
    \label{eq:axionEOMUV}
    a^{\prime \prime}
    -
    \frac{1}{m_\phi^2 R^2}
    \nabla^2
    a
    +
    \frac{
        3H
    }{
        m_\phi
    }
    a^{\prime}
    +
    \frac{
        f_a
        \,
        l(\phi)
    }{
        1
        +
        2
        f(\phi)
    }
    \sin (a / f_a)
    \\
    =
    -
    \frac{
        2
    }{
        1
        +
        2
        f(\phi)
    }
    \frac{
        g(\phi)
    }{
        m_P
    }
    \left[ 
        \phi^{\prime}
        a^{\prime}
        -
        \frac{1}{m_\phi^2 R^2}
        \nabla
        \phi 
        \cdot 
        \nabla
        a
    \right]
\end{multline}
where again we use primes to denote differentiation with respect to $z \equiv m_\phi t$, and we have defined the contributions from the axion kinetic couplings in terms of the dimensionless functions
\begin{equation}
    f(\phi) 
    \equiv
    \sum 
    \limits_{n=1}^\infty 
    c_n
    \left(
        \frac{\phi}{m_P}
    \right)^n
    \qquad
    \text{and}
    \qquad
    g(\phi) 
    \equiv
    \sum 
    \limits_{n=1}^\infty 
    n
    c_n
    \left(
        \frac{\phi}{m_P}
    \right)^{n-1}
\end{equation}
and contributions from the axion potential couplings in terms of the dimensionless functions
\begin{equation}
    l(\phi)
    \equiv
    \sum 
    \limits_{n=0}^\infty 
    \frac{d_n}{m_\phi^2 f_a^2}
    \left(
        \frac{\phi}{m_P}
    \right)^n
    \qquad 
    \text{and}
    \qquad 
    h(\phi)
    \equiv
    \sum 
    \limits_{n=1}^\infty 
        \frac{n \, d_n}{m_\phi^2 f_a^2}
    \left(
        \frac{\phi}{m_P}
    \right)^{n-1}
    .
\end{equation}

Throughout this section, we incorporate a few benchmark cases for the coupling coefficients\footnote{It is interesting to consider analyticity constraints on these coefficients where a well defined S-matrix in non-gravitational theories may determine the sign of the coefficients, however it has been noted that in quantum gravity there are known examples that violate these conditions and yet give consistent theories \cite{Adams:2006sv}. In this work, we do not impose any constraint on the coefficient sign and simply regard them as free parameters.}  $c_n$ (where $n\geq 1$) in terms of a unified parameter $\xi$,
\begin{equation}
    \label{eq:benchmarkCases}
    c_n
    \equiv
    \begin{cases}
        (-\xi)^n / (h\, n!)
        &
        \quad 
        \text{Benchmark I}
        \\
        -\xi/2
        &
        \quad
        \text{Benchmark II}
        \\
        (-1)^n \xi/2
        &
        \quad 
        \text{Benchmark III,}
    \end{cases}
\end{equation}
where Benchmark I is motivated by general dilatonic couplings, and Benchmarks II and III respectively reflect higher dimensional operators contributing ``constructively'' (same-sign for $\phi > 0$) or ``destructively'' (alternating-sign for $\phi > 0$). 
We also note that $f(\phi)$ for Benchmarks II and III are related through the simultaneous interchange $\phi \rightarrow - \phi$ and $\xi \rightarrow - \xi$, and $g(\phi)$ for Benchmarks II and III are related through the interchange $\phi \rightarrow -\phi$.
Thus, Benchmarks II and III may both exhibit constructive or destructive properties depending on the signs of $\phi$ and $\xi$.

\subsection{Modulus dynamics from higher-order operators}
We begin by analyzing the modulus back-reaction in \eqref{eq:modulusEOMUV}.
As in the case where we retained only the $n=1$ term, the back-reaction on the modulus generally appears to be negligible at first glance due to the separation of scales $\partial_\mu a \lesssim a \sim f_a \lesssim \phi \sim m_P$.
However, a large back-reaction could still be possible if $g(\phi)$ or $h(\phi)$ converges to a large value -- i.e. depending on the couplings $c_n$, we could have $g(\phi) \gtrsim m_P^2/f_a^2$ despite $\phi < m_P$.
To investigate this possibility further, we first focus on $g(\phi)$ in the Benchmark II scenario -- where higher-order terms are constructive -- and take\footnote{From these results, we would then likewise expect scenarios with $c_{n+1} < c_n$ and $c_{n+1} \sim - c_n$ to have higher-order contributions to the modulus back-reaction even further suppressed.}
$c_n = 1$.
Under this assumption, the series can then be straightforwardly evaluated.
For the contribution from $g(\phi)$ to be potentially sizable, we would expect the condition 
\begin{equation}
    \label{eq:modulusBackreactionCondition1}
    \frac{
        g(\phi \sim \phi_0)        
    }{
        m_P
    }
    a_0^2
    =
    \frac{
        a_0^2 / m_P
    }{
        (
            \phi_0/m_P
            -
            1
        )^2 
    }
    \simeq
    \phi_0
    .
\end{equation}
From this condition, we then see that for $a_0 / \phi_0 \sim 0.1$, the back-reaction contribution is only $\mathcal{O}(1)$ when $\phi_0 / m_P \gtrsim 0.90$. 
Even for the case when the axion has a Planck-scale amplitude ($a_0 \sim \phi_0 \sim m_P$), a large back-reaction contribution requires $\phi_0 / m_P \gtrsim 0.38$.
\begin{figure}[htb!]
    \centering
    \includegraphics[scale=0.4]{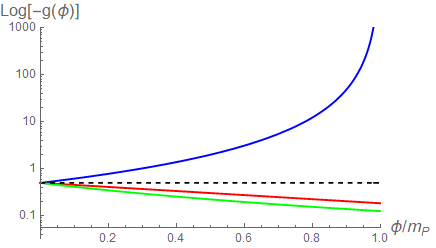}
    \includegraphics[scale=0.4]{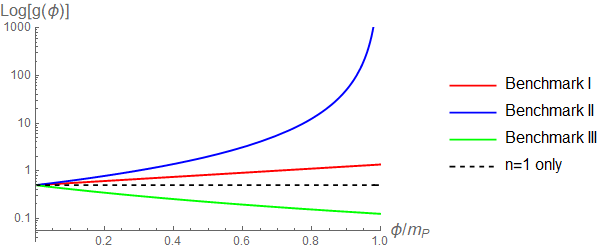}
    \caption{Plots of the size of the back-reaction prefactor $g(\phi)$ for our three benchmark scenarios. Benchmark I (red) is motivated by dilatonic couplings, while Benchmarks II (blue) and III (green) have higher-order terms interfere ``constructively'' and ``destructively,'' respectively. We also include the $n=1$ contribution only (black) for comparison with our toy model analysis. We take $\xi=1$ (left) and $\xi=-1$ (right) for illustration.}
    \label{fig:back-reactionCouplingValues}
\end{figure}

In Figure~\ref{fig:back-reactionCouplingValues}, we show the magnitude of $g(\phi)$ in each of our three benchmark cases.
It is evident that a large degree of constructive interference must occur for $g(\phi)$ to become large, in addition to a large modulus misalignment $\phi_0 \sim m_P$.
A large destructive interference, on the other hand, drives $g(\phi) \rightarrow 0$ and thus can heavily reduce or remove any potential back-reaction.

We can likewise study the back-reaction contribution from $h(\phi)$.
Under the assumption that each $d_n \sim m_a^2 f_a^2$, we see that the $n$-th term is expected to have a prefactor of order $m_a^2 / m_\phi^2$.
Under these conditions, we would expect the back-reaction contribution from $h(\phi)$ to become sizable if 
\begin{equation}
    \label{eq:modulusBackreactionCondition2}
    \frac{
        h(\phi \sim \phi_0)    
    }{
        m_P
    }
    f_a^2
    =
    \frac{m_a^2}{m_\phi^2}
    \frac{
        f_a^2/m_P
    }{(\phi_0/m_P-1)^2}
    \simeq
    \phi_0
    .
\end{equation}
Comparing \eqref{eq:modulusBackreactionCondition2} with \eqref{eq:modulusBackreactionCondition1} and assuming $a_0 \sim f_a$, we see that the contributions from $h(\phi)$ are expected to be far subdominant to those from $g(\phi)$ given that we expect $m_a \ll m_\phi$.
Thus, the addition of higher-order correction terms should not significantly alter our expectation that the modulus back-reaction is typically negligible and thus is largely unimportant to the modulus oscillations, with the exception being a regime in which significant axion growth through the preheating process occurs.
This can also be seen explicitly by recasting \eqref{eq:modulusEOMUV} in terms of the redefined field, $\Phi \equiv \phi R^{3/2}$:
\begin{multline}
    \label{eq:modulusDecayUV}
    \Phi^{\prime \prime}_k
    +
    \Bigg(
        \frac{\tilde{k}_\phi^2}{m_\phi^2}
        -
        \frac{9H^2}{4 m_\phi^2}
        +
        1
        -
        \phi^{-1}
        \left[ 
            \frac{
                f_a^2
                \,
                h(\phi)            
            }{
                m_P
            }
            \cos(a/f_a)
            +
            \frac{
                g(\phi)            
            }{
                m_P
            }
            \left[
                (a^{\prime})^2
                -
                \frac{1}{m_\phi^2 R^2}
                \left(
                    \nabla
                    a
                \right)^2
            \right]
        \right]
    \Bigg)
    \Phi_k
    = 
    0
    .
\end{multline}
From the expected hierarchy between $g(\phi)$ and $h(\phi)$ and the form of the potential and kinetic couplings, we see that the kinetic coupling has a much larger impact on the modulus instability.
We also arrive at the conclusion that, if the axion modes grow sufficiently that the condition 
$
g(\phi) (\partial_\mu a)^2 \gtrsim m_P^2
$ 
can be fulfilled, an exponential decay in the homogeneous modulus mode is possible.
However, if this condition cannot be fulfilled, the modulus dynamics effectively reduce to that of a free scalar field.

\subsection{Axion dynamics from higher-order operators}
We now proceed to study the axion dynamics in the presence of the higher-order terms.
By inspection of \eqref{eq:axionEOMUV}, we see a tachyonic instability is clearly present if the condition 
\begin{equation}
    f(\phi)
    <
    -
    \frac{1}{2}
\end{equation}
is satisfied.
In each benchmark case, the minimum value of $\phi_0 / m_P$ required to drive the axion mass tachyonic through this condition can be be found to be
\begin{equation}
    \phi_0/m_P
    \simeq
    \begin{cases}
        \xi^{-1}
        \log(2 / (2-h))
        &
        \quad 
        \text{Benchmark I}
        \\
        (\xi+1)^{-1}
        &
        \quad
        \text{Benchmark II}
        \\
        (\xi-1)^{-1}
        &
        \quad 
        \text{Benchmark III}
    \end{cases}
\end{equation}
where, in the first case a canonically normalized axion imposes $h=2$, while the last case requires $\phi_0/m_P < 0$ for $\xi<0$ (and thus can be identified with our constructive benchmark in this regime).
We are led to the conclusion that UV couplings of dilatonic form
\begin{equation}
    S
    \supset
    \int 
    d^4x
    \sqrt{-g}
    \left(
        \frac{1}{2}
        e^{-\xi \phi/m_P}
        \,
        \partial_\mu 
        a
        \,
        \partial^\mu 
        a
    \right)
\end{equation}
appear to be free of this particular instability, which would require $\phi_0/m_P \rightarrow \infty$ -- at least barring any additional effects.
\begin{figure}[htb!]
    \centering
    \includegraphics[scale=0.4]{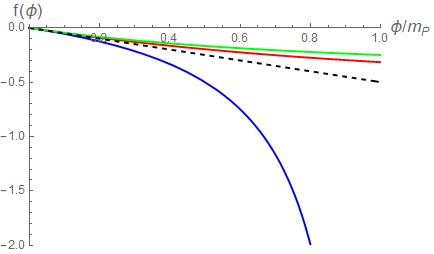}
    \includegraphics[scale=0.4]{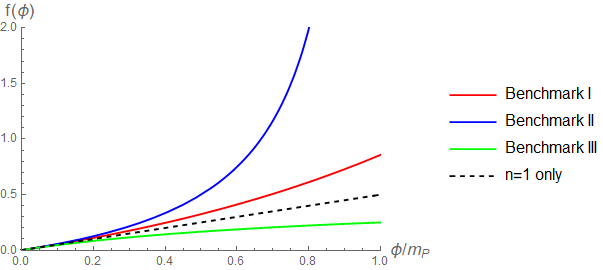}
    \caption{Plots of the size of the kinetic coupling $f(\phi)$ for our three benchmark scenarios. Benchmark I (red) is motivated by dilatonic couplings, while Benchmarks II (blue) and III (green) have higher-order terms interfere ``constructively'' and ``destructively,'' respectively. We also include the $n=1$ contribution only (black) for comparison with our toy model analysis. We take $\xi=1$ (left) and $\xi=-1$ (right) for illustration.}
    \label{fig:kineticCouplingValues}
\end{figure}
In Figure~\ref{fig:kineticCouplingValues}, we illustrate the size of the kinetic coupling, $f(\phi)$, for each of these benchmarks as well as for the dimension-five term only, which coincides for all three benchmark cases.

Moving to the terms in \eqref{eq:axionEOMUV} which arise due to the modulus coupling to the axion potential, we see that the presence of $l(\phi)$ largely serves only as a modification to the axion effective mass and thus has a quantitative impact on the growth rate for any instability.
However, under the assumption that $d_n \sim m_a^2 f_a^2$ we expect $l(\phi) \sim \mathcal{O}(m_a^2 /m_\phi^2) \ll 1$.
Unless $\phi / m_P$ is tuned sufficiently close to 1, we do not anticipate the presence of $l(\phi)$ to significantly affect our expectations from our analysis with the $n=0$ term only -- and hence why we have dropped the dimension-five operator $(d_1/m_P) \, \phi \, m_a^2 a^2$ in our toy model.  

In addition to the tower of modulus couplings, anharmonic contributions due to the axion potential are present.
If the axion satisfies $a/f_a \ll 1$ (i.e. its initial position is not at the top of its true potential), classical perturbation theory indicates that these anharmonic terms can be expected to provide small shifts to the effective mass term -- although the qualitative behavior is again unchanged.
However, if the axion has $a \sim f_a$ the equation of motion becomes highly non-linear and thus may induce new instabilities. 
Full treatment of this case goes beyond the scope of this work, however we expect that this may lead to rich phenomenology.

Finally, we move to the additional friction term for the axion and note that its behavior is primarily dictated by $g(\phi)$.
However, since we have $\phi \propto z^{-1}$ during matter domination and $\phi \propto z^{-3/4}$ during radiation domination, we expect the additional contributions from $\phi^{n-1}$ that are embedded in $g(\phi)$ to rapidly become negligible.
At very early times (i.e. $z\ll 1$), the additional axion friction term is also likely to be negligible, as we expect $\phi^\prime \sim 0$.
Thus, it appears that the axion dynamics are effectively unchanged from the minimally-coupled scenario outside of a regime where significant preheating-like effects occur.

We now shift to investigating the regime where the modulus has begun to oscillate in a more quantitative sense, and study the implications on the axion stability. 
Under the field redefinition 
$
    A_k
    \equiv 
    a_k
    R^{3/2}
    \left(
        1
        +
        2
        f(\phi)
    \right)^{1/2}
$, 
the axion equation of motion \eqref{eq:axionEOMUV} can be written as
\begin{equation}
    \label{eq:generalizedAxionEOM}
    A_k^{\prime \prime}
    +
    \Bigg(
        -
        \left(
            \frac{
                \frac{
                    g^\prime(\phi)
                }{
                    m_P^2
                }
                \phi^\prime
                +
                \frac{
                    g(\phi)
                }{
                    m_P^2
                }
                \phi^{\prime \prime}
            }{
                1
                +
                2
                f(\phi)
            }
        \right)
        +
        \left(
            \frac{
                \frac{
                    g(\phi)
                }{
                    m_P
                }
                \phi^\prime
            }{
                1
                +
                2
                f(\phi)
            }
        \right)^2
        +
        \frac{\tilde{k}^2}{m_\phi^2}
    \Bigg)
    A_k
    =
    0
\end{equation}
where we have again neglected $H^\prime$ terms and take the limit where $H/m_\phi \ll 1$, and we have also neglected the $l(\phi)$ term as we generically expect $l(\phi) \ll 1$.
Applying this result to the late-time regime, i.e. where we may assume
\begin{equation}
    \phi \simeq \phi_0 \cos(z),
\end{equation}
a few general features are apparent.
Perhaps unsurprisingly, $f(\phi)$ and $g(\phi)$ are periodic in this regime.
\eqref{eq:generalizedAxionEOM} is thus a form of Hill's equation, regardless of the values of $c_n$.
This leads us to the expectation that UV theories which produce couplings between axionic kinetic terms and moduli may generically contain regimes of instability and hence warrant extensive opportunity for future study.
To illustrate this point, we investigate the presence of instabilities in \eqref{eq:generalizedAxionEOM} for each of our three benchmark cases.  
We also make the convenient definition 
\begin{equation}
    \alpha 
    \equiv \phi_0 / m_P
\end{equation}
throughout to simplify notation.

\subsubsection{Benchmark I: the dilatonic case}
For the general case of a dilatonic coupling
$
    c_n
    =
    (-\xi)^n / (2n!),
$
the series $f(\phi)$ and $g(\phi)$ become
\begin{equation}
    f(\phi)
    =
    \frac{1}{2}
    \left(
        -1
        +
        e^{
            -
            \alpha
            \xi
            \cos(z)
        }
    \right)
    \qquad 
    \text{and}
    \qquad
    g(\phi)
    =
    -
    \frac{\xi}{2}
    \,
    e^{
        -
        \alpha
        \xi 
        \cos(z)
    }
\end{equation}
in the late-time limit.
This produces the axion equation of motion
\begin{equation}
    \label{eq:stabilityEomUvDilaton}
    A_k^{\prime \prime}
    +
    \Bigg(
        -
        \frac{\alpha \xi}{2}
        \cos(z)
        -
        \frac{\alpha^2 \xi^2}{4}
        \sin^2(z)
        +
        \frac{\tilde{k}^2}{m_\phi^2}
    \Bigg)
    A_k
    =
    0
    .
\end{equation}
We see that for $\alpha \xi \ll 1$, we arrive at the same frequency given by \eqref{eq:wkbFrequencyLateTimes} for the lone dimension-five operator, with the identification 
$
    -
    \alpha \xi
    =
    2
    \beta
$.

\begin{figure}[htb!]
    \centering
    \includegraphics[width=0.49\linewidth]{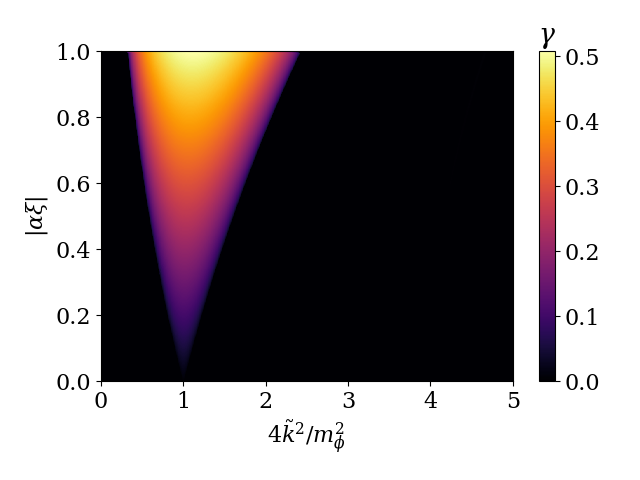}
    \includegraphics[width=0.49\linewidth]{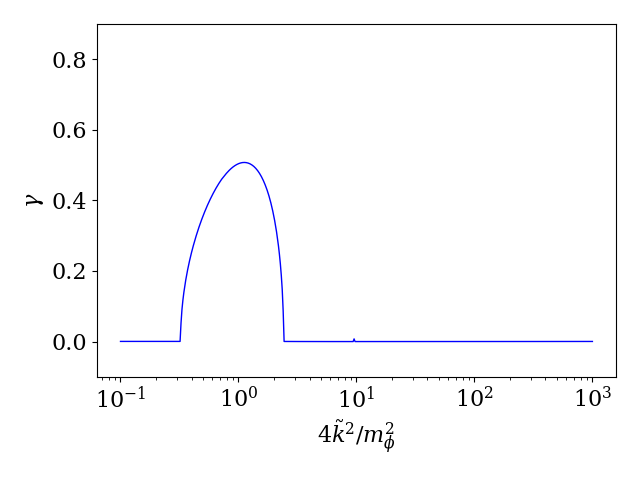}
    \caption{Left: axion growth rate for Benchmark I (the dilatonic case) for the $\mathcal{A}_k \equiv 4 \tilde{k}^2/m_\phi^2 \leq 5$ and $|\alpha \xi| \equiv |\xi \phi_0 / m_P| \leq 1$ domains.
    Right: growth rate for the particular choice of $|\alpha\xi|=1$.}
    \label{fig:stabilityUvDilaton}
\end{figure}

We display the growth rate, $\gamma$, for the axion in Figure~\ref{fig:stabilityUvDilaton}. 
A few features are directly evident when we compare this with the dimension-five case in Figure~\ref{fig:mathieu-late-times}.
Although there are slight quantitative changes, we see that the overall qualitative behavior for the axion growth is largely unchanged -- modes with $\tilde{k} \sim m_\phi / 2$ result in growth with a rate that increases for stronger couplings, $\xi$, and larger initial field amplitudes, $\alpha$.
We also notice a similar quantitative value of the growth rate between the two cases, peaking at around $\gamma \sim 0.5$ within the parameter space considered. 
However, the full dilatonic case shown in Figure~\ref{fig:stabilityUvDilaton} suggests that bands of $\tilde{k}/m_\phi$ which result in instability are pushed towards slightly higher momenta for this case -- i.e. the higher-order terms serve to excite smaller axion wavelengths.
This shift is subtle enough that we do not necessarily expect a modulus decay efficiency drastically different from the dimension-five case we studied previously, as can be seen from \eqref{eq:modulusDecayUV} and noting the similarity in $g(\phi)$ values.

\begin{figure}[htb!]
    \centering
    \includegraphics[scale=0.6]{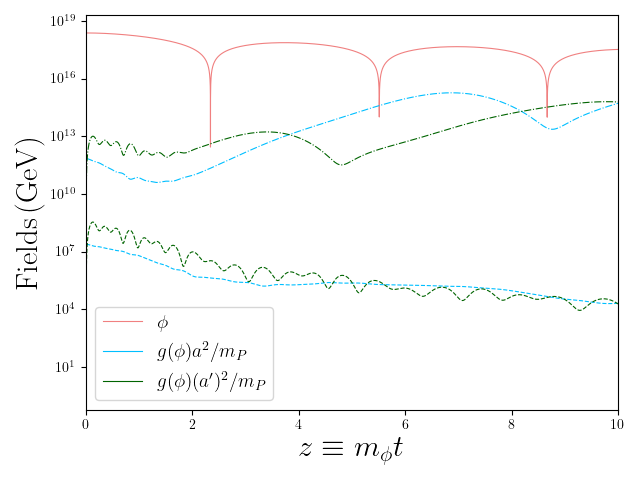}
    \caption{Magnitudes of the modulus field (red) and the back-reaction due to the axion field (blue) and axion velocity (green).
    Dashed curves have $\alpha \xi = -2.3$ and dot-dashed curves have $\alpha \xi = -11$. Here, we take $a_0/\phi_0 = 10^{-1}$.}
    \label{fig:dilatonFieldsLargeXi}
\end{figure}

In Figure~\ref{fig:dilatonFieldsLargeXi}, we show the back-reaction contributions to the modulus from the axion field and axion velocity, which are numerically computed from \eqref{eq:modulusEOMUV}-\eqref{eq:axionEOMUV} for the dilatonic case following our procedure from Section~\ref{AMD}.
Here, the dashed curves correspond to $\alpha \xi = -2.3$, which matches our choice of $c\phi_0 / m_P = 1.2$ for the dimension-five term only in Figure~\ref{fig:dim5FieldsLargeXi}.
Comparing these two figures, we see similar behavior and magnitudes for the back-reactions shown by the dashed curves, as expected, so that the enhancement of the axion field for small (i.e. $\lesssim \mathcal{O}(1)$) couplings and field amplitudes is not drastically changed once the additional corrections are taken into account.
However, as we now consider a UV theory that is free of the naive $f(\phi)<-1/2$ tachyonic instability, we consider a much larger value of the coupling ($\alpha \xi = -11$) shown by the dot-dashed curves in Figure~\ref{fig:dilatonFieldsLargeXi}.
Such a large coupling enhances the axion field enough that it \textit{can} produce a sizable back-reaction, and signifies the potential breakdown of our approximations in our numerical treatment.

\begin{figure}[htb!]
    \centering
    \includegraphics[scale=0.47]{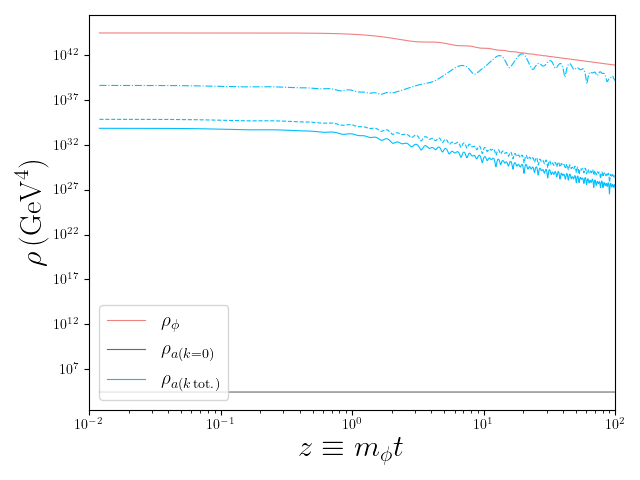}
    \includegraphics[scale=0.47]{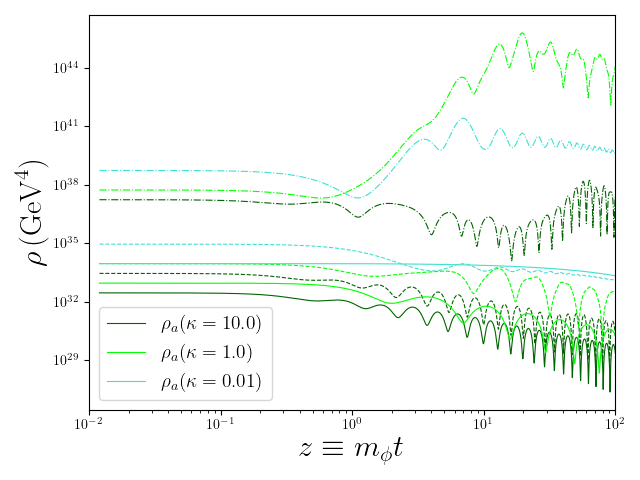}
    \caption{Left: Energy densities of the modulus (red) and axion (blue) fields. We also display the axion zero mode (gray) for comparison. Right: Contributions to the axion energy density from individual modes.
    We define $\kappa \equiv 4\tilde{k}^2 / m_\phi^2$, where $\tilde{k}$ is taken to be the physical momentum at the onset of modulus oscillations. 
    In both plots, solid curves denote the absence of gravitational coupling, while dashed curves have $\alpha \xi = -2.3$ and dot-dashed curves have $\alpha \xi = -11$.
    Both plots take $a_0 /\phi_0 = 10^{-1}$.}
    \label{fig:dilatonEnergyLargeXi}
\end{figure}

We show the average energy density of the axion field (blue) and the modulus (red) in the left panel of Figure~\ref{fig:dilatonEnergyLargeXi}.
It is immediately evident that for $\alpha \xi = -11$, as shown by the blue dot-dashed curve, the axion field can be enhanced to a level comparable to the modulus field. 
Although our numerical treatment here does not factor in the back-reaction to the modulus -- nor the production and rescattering of non-zero moduli modes -- we might expect a relatively high decay efficiency from such large coupling values, however it still remains to be seen if the near-perfect efficiency required to solve the cosmological moduli problem can be attained here.
It is worth noting that in \cite{Adshead:2024ykw}, a similar model was studied in an inflationary reheating context and a high efficiency ($\sim 10^{-1}$) was produced for such a coupling value, although we leave it for future work to investigate the efficiency of our post-inflationary case through the use of lattice simulations \cite{upcomingWork}.
On the right panel of Figure~\ref{fig:dilatonEnergyLargeXi}, we display the contributions to the energy density from individual modes which have $\kappa \equiv 4\tilde{k}^2 / m_\phi^2 = 0.01$ (blue), $1$ (light green), and $10$ (dark green), where $\tilde{k}$ is again taken to be the physical momentum at the onset of modulus oscillations.
We again see pronounced amplification of modes with $4\tilde{k}^2 / m_\phi^2 \sim 1$ as the gravitational coupling is increased, while super-horizon modes receive a much smaller amplification once they enter the Hubble radius, and modes which are sub-horizon at the onset of moduli oscillations must redshift before receiving an albeit smaller amplification relative to those with $4 \tilde{k}^2/m_\phi^2 = 1$ at the onset of moduli oscillations.

\begin{figure}[htb!]
    \centering
    \includegraphics[scale=0.6]{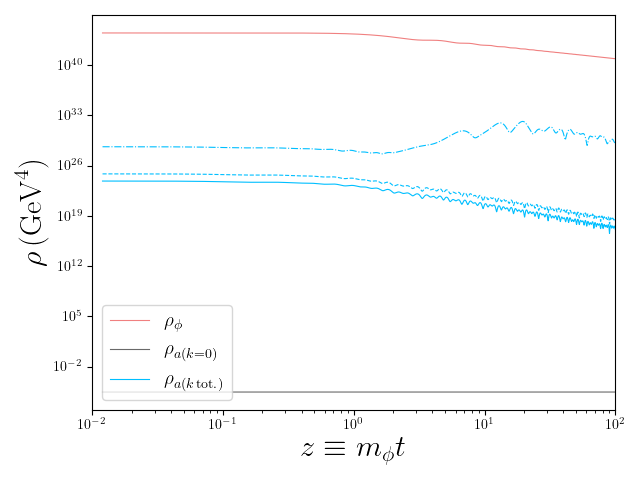}
    \caption{Energy densities of the modulus (red) and axion (blue) fields. We also display the axion zero mode (gray) for comparison. 
    Solid curves denote the absence of gravitational coupling, while dashed curves have $\alpha \xi = -2.3$ and dot-dashed curves have $\alpha \xi = -11$.
    Here, we take $a_0 /\phi_0 = 10^{-6}$.}
    \label{fig:dilatonEnergySmallXi}
\end{figure}

In Figure~\ref{fig:dilatonEnergySmallXi}, we display the energy densities assuming instead a much lower value of the PQ scale, so that $a_0 \sim f_a \sim 10^{12}$ GeV -- a value consistent with expectations for the QCD axion.
As this value correspondingly lowers the initial condition for the non-zero axion modes, we see the axion field amplification is well below the threshold to significantly impact the evolution of the modulus.
Thus, it appears that a large PQ scale is necessary -- although potentially not sufficient -- to fully decay the modulus through this non-perturbative effect.
Additionally, although lower PQ scales may enhance the axion population and behave as either dark radiation or dark matter, the late perturbative decay will inject sufficient entropy to render this enhancement negligible unless such a high level of efficiency is achieved.
However, this does present the possibility of enhanced axion perturbations which may persist -- and later collapse and form axion substructures. 

\subsubsection{Benchmark II: constructive higher-order terms}
For the benchmark case where $c_n=-\xi/2$, i.e. where higher-order terms are constructive for $\phi > 0$, we have the dimensionless functions
\begin{equation}
    f(\phi)
    =
    \frac{\xi}{2}
    \frac{
        \alpha 
        \cos (z)
    }{
        \alpha 
        \cos (z)
        -
        1
    }
    \qquad 
    \text{and}
    \qquad 
    g(\phi)
    =
    -
    \frac{\xi}{2}
    \frac{
        1
    }{
        ( 
            \alpha 
            \cos (z)
            -
            1
        )^2
    }
    .
\end{equation}
These then produce the axion equation of motion
\begin{multline}
    A_k^{\prime \prime}
    +
    \Bigg(
        \frac{\xi}{2}
        \frac{
            \alpha 
            \cos(z)
            -
            \alpha^2
            \sin^2(z)
            -
            \alpha^2
        }{
            \left(
                \alpha
                (
                    \xi
                    +
                    1
                )
                \cos(z)
                -
                1
            \right)
            (\alpha \cos(z)-1)^2
        }
        \\
        +
        \frac{\xi^2}{4}
        \frac{
            \alpha^2
            \sin^2(z)
        }{
            \left(
                \alpha
                (
                    \xi
                    +
                    1
                )
                \cos(z)
                -
                1
            \right)^2
            (\alpha \cos(z)-1)^4
        }
        +
        \frac{\tilde{k}^2}{m_\phi^2}
    \Bigg)
    A_k
    =
    0
    .
\end{multline}
In the limit where both $\alpha \ll 1$ and $\xi \ll 1$, we again arrive at the same axion frequency as for the dimension-five term only, \eqref{eq:wkbFrequencyLateTimes}, with the identification $-\alpha \xi = 2\beta$.

\begin{figure}[htb!]
    \centering
    \includegraphics[width=0.49\linewidth]{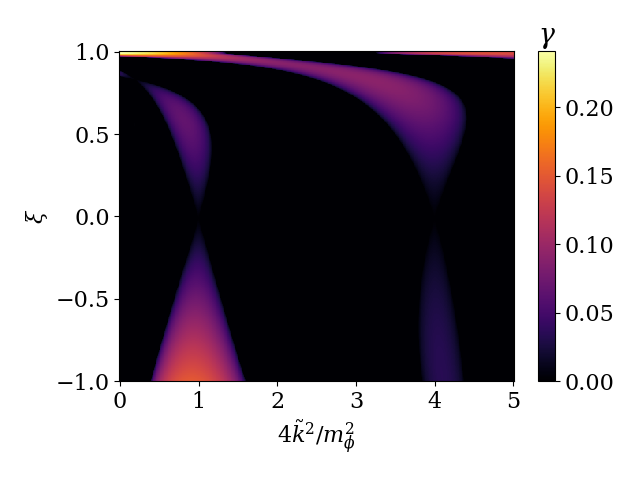}
    \includegraphics[width=0.49\linewidth]{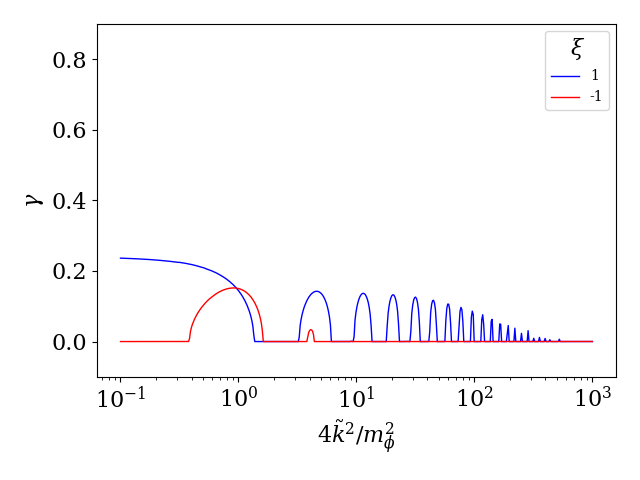}
    \caption{Left: growth rate for Benchmark II (constructive UV terms) for the $\mathcal{A}_k \equiv 4 \tilde{k}^2/m_\phi^2 \leq 5$ and $-1 \leq \xi \leq 1$ domains. Here, we fix $\alpha \equiv \phi_0 / m_P = 0.49$ to isolate the instability bands of Hill's equation.
    Right: growth rate for the particular choice of $\xi=\pm1$.}
    \label{fig:stabilityUvConstructive}
\end{figure}

It is interesting to note that in this benchmark case, the $f(\phi)=-1/2$ instability can be present for $0<\xi<1$ if $\alpha \geq 1/2$.
However, this instability is not present for $-1<\xi<0$ (even if $1/2\leq \alpha \leq 1$) so that any instability in this range arises solely from the bands of the Hill equation.
We show the growth rate $\gamma$ of the axion for $\alpha = 0.49$ in Figure~\ref{fig:stabilityUvConstructive}, which ensures the instability from $f(\phi)=-1/2$ is absent and thus the validity of our field redefinition to $A_k$.
Compared to Benchmark I (Figure~\ref{fig:stabilityUvDilaton}) and the lone dimension-five contribution (Figure~\ref{fig:mathieu-late-times}), we see the appearance of new bands of instability for $\xi>0$, while $\xi<0$ appears similar to the former cases.
In the case of $\xi > 0$, we have a large enhancement to the effective potential of $A_k$ through $(1+2f(\phi))^{-1}$, resulting in a large increase in the energy of the produced axions which is reflected by the resulting instability bands at $4\tilde{k}^2/m_\phi^2 \gg 1$.
Thus, we might expect Benchmark II with $\xi>0$ to enhance the efficiency of the modulus decay due to the excitement of the small-wavelength modes, although we again leave detailed study of this case for future work.
For the case where $\xi<0$, there is no enhancement to the effective potential of $A_k$, and we once again only see significant instability if $\tilde{k} \sim m_\phi / 2$.

\subsubsection{Benchmark III: destructive higher-order terms}
For the benchmark case where $c_n=(-1)^n\xi / 2$, we have
\begin{equation}
    f(\phi)
    =
    -
    \frac{\xi}{2}
    \frac{
        \alpha \cos (z)
    }{
        \alpha \cos (z)
        +
        1
    }
    \qquad
    \text{and}
    \qquad
    g(\phi)
    =
    -
    \frac{\xi}{2}
    \frac{
        1
    }{
        (
            \alpha \cos (z)
            +
            1
        )^2
    }
\end{equation}
which produces the axion equation of motion
\begin{multline}
    A_k^{\prime \prime}
    +
    \Bigg(
        \frac{\xi}{2}
        \frac{
            \alpha
            \cos(z)
            +
            \alpha^2
            \sin^2(z)
            +
            \alpha^2
        }{
            \left(
                \alpha 
                (
                    \xi
                    -
                    1
                )
                \cos(z)
                +
                1
            \right)
            (
                \alpha
                \cos(z)
                +
                1
            )^2
        }
        \\
        +
        \frac{\xi^2}{4}
        \frac{
            \alpha^2
            \sin^2(z)
        }{
            \left(
                \alpha 
                (
                    \xi
                    -
                    1
                )
                \cos(z)
                +
                1
            \right)^2
            (
                \alpha
                \cos(z)
                +
                1
            )^4
        }
        +
        \frac{\tilde{k}^2}{m_\phi^2}
    \Bigg)
    A_k
    =
    0
    .
\end{multline}
As with both the Benchmark I and II cases, the $\alpha \ll 1$ and $\xi \ll 1$ limit reproduces the lone dimension-five frequency given by \eqref{eq:wkbFrequencyLateTimes}, with the identification $-\alpha \xi = 2 \beta$.

\begin{figure}[htb!]
    \centering
    \includegraphics[width=0.49\linewidth]{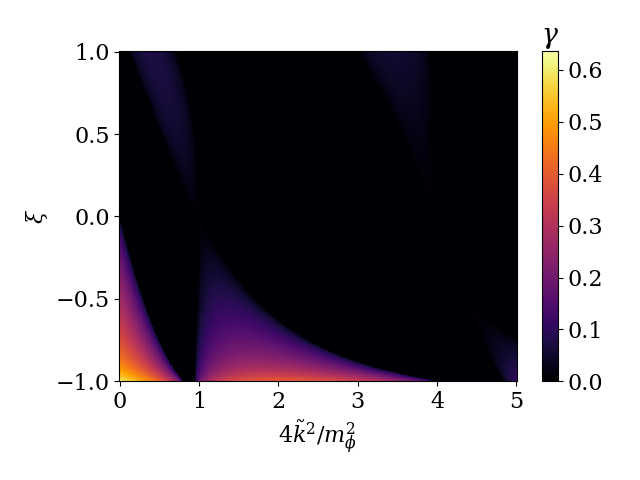}
    \includegraphics[width=0.49\linewidth]{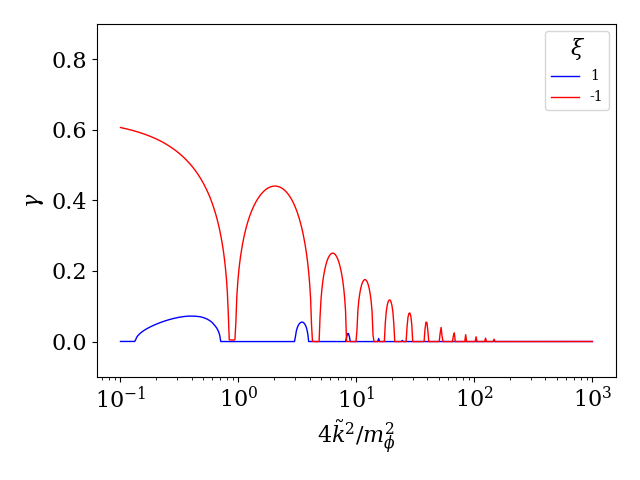}
    \caption{Left: growth rate for Benchmark III (destructive UV terms) for the $\mathcal{A}_k \equiv 4 \tilde{k}^2/m_\phi^2 \leq 5$ and $-1 \leq \xi \leq 1$ domains. Here, we fix $\alpha \equiv \phi_0 / m_P = 0.49$ to isolate the instability bands of Hill's equation.
    Right: growth rate for the particular choice of $\xi=\pm1$.}
    \label{fig:stabilityUvDestructive}
\end{figure}

Similarly to the Benchmark II case, the $f(\phi) = -1/2$ instability can be present for $-1 < \xi < 0$ if $\alpha \geq 1/2$. 
In Figure~\ref{fig:stabilityUvDestructive}, we show the growth rate, $\gamma$, of the axion for $\alpha = 0.49$ which again ensures the absence of the $f(\phi)=-1/2$ instability.
It is evident that for $\xi>0$, any axion growth is highly suppressed compared to the dimension-five only case -- and thus we expect any preheating effects to be excluded in this regime.
For $\xi<0$, we again see the appearance of additional bands of instability. 
Here, the effective potential of $A_k$ can be enhanced through $(1+2f(\phi))^{-1}$ similarly to the Benchmark II case, where again $f(\phi)$ for Benchmarks II and III can be related through the interchange of $\xi \rightarrow -\xi$ and $\phi \rightarrow -\phi$.
However, $g(\phi)$ differs from the corresponding Benchmark II case by a sign -- causing the instability band to occur for lower $\tilde{k}/m_\phi$ by comparison.
Again, detailed study of this case is left for future work; however, we might expect that this benchmark case with $-1 < \xi < 0$ yields efficient preheating, although the shift in axion modes will likely also affect the results in a quantitative sense.

\subsection{Application to Large Volume Scenario Models}
We now discuss the previous results within the context of the Large Volume Scenario (LVS) models in string theory.
LVS models are characterized by compactifications where the volume of the compactification manifold is well-approximated by one or two ``bulk'' K\"ahler moduli, and potentially many ``blow-up'' K\"ahler moduli which are extremely small by comparison.
Moduli stabilization is then achieved 
by balancing non-perturbative corrections to the superpotential (due to e.g. $\overline{D}3$-branes, instanton corrections, etc.) against perturbative $\alpha'$ corrections to the K\"ahler potential. 
One particular class of LVS models, known as fibered LVS due to the structure of the compactification volume, also requires the inclusion of perturbative $g_s$ corrections to stabilize one of the two bulk moduli.
We refer the reader to \cite{Balasubramanian:2005zx,Conlon:2005ki,Cicoli:2012aq} for additional details on the class of LVS models with one bulk modulus (minimal LVS) and \cite{Cicoli:2007xp,Cicoli:2008va,Cicoli:2018cgu} for additional details on the class of LVS models with two bulk moduli (fibered LVS).

In the limit of large volume, the relevant K\"ahler potential assumes the form 
\begin{equation}
    K 
    \simeq
    -2
    \log \mathcal{V}
\end{equation}
where the volume can be written in terms of the four-cycle moduli ($\tau_b$ for minimal LVS and $\tau_{1,2}$ for fibered LVS) as 
\begin{equation}
    \mathcal{V}
    \simeq
    \begin{cases}
        \tau_b^{3/2} 
        &
        \text{(minimal LVS)}
        \\
        \sqrt{\tau_1}
        \tau_2
        &
        \text{(fibered LVS).}
    \end{cases}
\end{equation}
The corresponding axions, $c_i$, which arise due to the dimensional reduction of the $C_4$ form field along the appropriate divisors, then become the complex component of the K\"ahler coordinates, $T_i \equiv \tau_i + i c_i$ (sometimes also referred to as K\"ahler moduli in the literature).
In this way, the K\"ahler potential can be rewritten as 
\begin{equation}
    K
    =
    \begin{cases}
        -
        3
        \log 
        \left(
            T_b
            +
            \overline{T}_b
        \right)
        &
        \text{(minimal LVS)}
        \\
        -
        \log 
        \left(
            T_1
            +
            \overline{T}_1
        \right)
        -
        2
        \log 
        \left(
            T_2
            +
            \overline{T}_2
        \right)
        &
        \text{(fibered LVS).}
    \end{cases}
\end{equation}

The Lagrangian is then given by 
$
    \mathcal{L}
    \supset
    K_{i \overline{\jmath}}
    \partial_\mu 
    T^i
    \partial^\mu 
    \overline{T}^{\overline{\jmath}}
$,
where $K_{i \overline{\jmath}} = \partial_i \partial_{\overline{\jmath}} K$ is the K\"ahler metric, and $i$, $\overline{\jmath}$ run over the fields present in the K\"ahler potential, $K$.
Applying this to the minimal LVS case, we arrive at 
\begin{equation}
    \mathcal{L}_{\text{minimal LVS}}
    \supset
    \frac{3}{4 \tau_b^2}
    \left(
        \partial_\mu 
        \tau_b 
        \partial^\mu 
        \tau_b
        +
        \partial_\mu 
        c
        \partial^\mu 
        c
    \right)
\end{equation}
while for the fibered LVS case, we have
\begin{equation}
    \mathcal{L}_{\text{fibered LVS}}
    \supset
    \frac{1}{4 \tau_1^2}
    \left(
        \partial_\mu 
        \tau_1
        \partial^\mu 
        \tau_1
        +
        \partial_\mu 
        c_1
        \partial^\mu 
        c_1
    \right)
    +
    \frac{1}{2 \tau_2^2}
    \left(
        \partial_\mu 
        \tau_2
        \partial^\mu 
        \tau_2
        +
        \partial_\mu 
        c_2
        \partial^\mu 
        c_2
    \right)
    .
\end{equation}
In both cases, canonical kinetic terms for both the modulus and the axion can be produced by a field redefinition of the form $\tau_i = \exp(\alpha_i \phi_i / m_P)$ where $\alpha_i$ are numeric constants, then expanding the fields around the (canonical) moduli VEVs, $\phi_i = \langle \phi_i \rangle + \delta \phi_i$, and finally rescaling the axion field $c_i = \alpha_i \exp( \alpha_i \langle \phi_i \rangle /m_P) a_i $.
For simplicity of notation, we identify the perturbations around the VEV as the canonically normalized fields $\phi_i$ and $a_i$, as any factors of the VEV have vanished at this level of analysis.
The Lagrangian can be rewritten as
\begin{equation}
    \mathcal{L}_{\text{minimal LVS}}
    \supset
    \frac{1}{2}
    \partial_\mu 
    \phi
    \partial^\mu 
    \phi
    +
    \frac{1}{2}
    \exp(-2 \sqrt{2/3} \phi )
    \partial_\mu 
    a
    \partial^\mu 
    a
\end{equation}
for the minimal LVS case, and 
\begin{multline}
    \mathcal{L}_{\text{fibered LVS}}
    \supset
    \frac{1}{2}
    \partial_\mu 
    \phi_1
    \partial^\mu 
    \phi_1
    +
    \frac{1}{2}
    \partial_\mu 
    \phi_2
    \partial^\mu 
    \phi_2
    \\
    +
    \frac{1}{2}
    \exp(-2 \sqrt{2} \phi_1 )
    \partial_\mu 
    a_1
    \partial^\mu 
    a_1
    +
    \frac{1}{2}
    \exp(-2 \phi_2 )
    \partial_\mu 
    a_2
    \partial^\mu 
    a_2
\end{multline}
for the fibered LVS case.
In the case of fibered LVS, however, there is one final subtlety -- the moduli $\phi_1$ and $\phi_2$ must be rotated to be mass eigenstates, $\phi_\mathcal{V}$ and $\phi_u$.
These rotations are specified in \cite{Cicoli:2022uqa} and produce the Lagrangian 
\begin{multline}
    \mathcal{L}_{\text{fibered LVS}}
    \supset
    \frac{1}{2}
    \partial_\mu 
    \phi_\mathcal{V}
    \partial^\mu 
    \phi_\mathcal{V}
    +
    \frac{1}{2}
    \partial_\mu 
    \phi_u
    \partial^\mu 
    \phi_u
    \\
    +
    \frac{1}{2}
    \exp
    \left(
        -2 \sqrt{2/3}
        \phi_\mathcal{V}
    \right)
    \partial_\mu 
    a_1
    \partial^\mu 
    a_1
    +
    \frac{1}{2}
    \exp
    \left(
        -2 
        \sqrt{2/3}
        \phi_\mathcal{V}
    \right)
    \partial_\mu 
    a_2
    \partial^\mu 
    a_2
    \\
    +
    \frac{1}{2}
    \exp
    \left(
        4
        \sqrt{1/3}
        \phi_u
    \right)
    \partial_\mu 
    a_1
    \partial^\mu 
    a_1
    +
    \frac{1}{2}
    \exp
    \left(
        -2 
        \sqrt{1/3}
        \phi_u
    \right)
    \partial_\mu 
    a_2
    \partial^\mu 
    a_2
\end{multline}
where the moduli mass eigenstates, $\phi_\vo$ and $\phi_u$, generically satisfy $m_{\phi_\mathcal{V}} \gtrsim m_{\phi_u}$.
From these Lagrangians, we can now apply our results from the dilatonic coupling case to determine if either LVS model may evade the cosmological moduli problem -- and by extension, the dark radiation overproduction problem -- through non-perturbative decay.

For the minimal LVS case, we can make the identifications (assuming a maximal initial displacement, $\phi_0/m_P = 1$)
\begin{equation}
    \xi_{\text{minimal LVS}}
    =
    -
    2\sqrt{2/3}
    \simeq
    -1.6,
    \qquad 
    \beta_{\text{minimal LVS}}
    =
    -
    \sqrt{2/3}
    \simeq
    -
    0.82
\end{equation}
to both the full dilatonic treatment and the dimension-five only treatment.
It is immediately evident that, were we to consider only the dimension-five contribution in our treatment, we would arrive at the conclusion that minimal LVS would produce a tachyonic ground state for the axion after inflation.
However, once the full UV theory is examined, this naive instability is lifted.
Additionally, when we consider the similar growth rates between the full dilatonic case and the lone dimension-five treatment, and recall minimum value of $\beta$ required to achieve any substantial modulus decay from the Benchmark I case, we do not expect the modulus to non-perturbatively decay into axions in the minimal LVS case.
Thus, perturbative decays -- and the standard picture of a modulus-dominated cosmology -- appear to dominate for minimal LVS.

Within the fibered LVS case, the lightest modulus $\phi_u$ contains couplings to two axions, $a_1$ and $a_2$, with the identifications (again assuming maximal initial displacement, $\phi_{u,\, 0} / m_P = 1$)
\begin{alignat}{2}
    &
    \xi_{\text{fibered LVS}}^{(1)}
    =
    4
    \sqrt{1/3}
    \simeq
    2.3
    ,
    \qquad
    &&
    \beta_{\text{fibered LVS}}^{(1)}
    =
    2\sqrt{1/3}
    \simeq
    1.2
    \\
    &
    \xi_{\text{fibered LVS}}^{(2)}
    =
    -2
    \sqrt{1/3}
    \simeq
    -1.2,
    \qquad
    &&
    \beta_{\text{fibered LVS}}^{(2)}
    =
    -\sqrt{1/3}
    \simeq
    -
    0.58
\end{alignat}
again for both the full dilatonic treatment and the dimension-five only treatment. 
As in the case of minimal LVS, a naive tachyonic instability appears in the lone dimension-five treatment for the $a_2$ axion -- an instability which vanishes once the full UV treatment is considered.
However, we see that for the values of $\xi$ and $\beta$ produced in fibered LVS, there is the potential that the $a_1$ axion can be appreciably enhanced.
Although this enhancement is likely insufficient\footnote{
Non-perturbative axion production from a string modulus oscillating around the minimum of its potential was also studied recently in \cite{Leedom:2024qgr} with particular focus on fibered LVS models. 
One point of difference between the two studies is that the coefficients $d_n$ multiplying the axion potential term in (\ref{eq:actionHigherOrderGravitational}) were taken to  be  $d_n \sim \mathcal{O}(m_a^2 f_a^2)$ where $m_a \ll m_\phi$ in our work, while they were taken to have dilatonic scaling with $m_a \sim m_\phi/2$ in the work of Leedom et al. -- akin to the coefficients $c_n$ in Benchmark I of our (\ref{eq:benchmarkCases}).
We have checked that assuming dilatonic scaling for $\frac{d_n}{m_\phi^2 f_a^2} \sim 1$ does not radically change our results.
However, a comparison of coupling values shows that $\frac{d_n}{m_\phi^2 f_a^2} \gg 1$ in the work of Leedom et al., such that the dominant production in their work is due to the potential coupling.
Additionally, we believe that an explicit comparison of the energy density of the produced axions with the energy density retained by the modulus field reveals that in both works, although axions may be produced in considerable quantities via kinetic or potential couplings, the produced energy density is likely not generically large enough to trigger a significant non-perturbative decay of the modulus.
Thus, we expect both sets of results to be compatible in this way.} to change the standard picture of a modulus-dominated cosmology, we are also led to the possibility of an interesting scenario in which axion perturbations are produced in additional quantities through the partial non-perturbative decay, and continue to grow during the modulus-dominated epoch -- leading to further enhancement of produced axion miniclusters than previously studied \cite{Nelson:2018via,Visinelli:2018wza,Blinov:2019jqc,WileyDeal:2023trg}, and an opportunity for future work.

\section{Conclusions}
In this paper, we explored the implications of a non-renormalizable coupling between moduli and axions on the dynamics of both fields in an expanding universe.
Such couplings are generically expected in string theory models, which typically produce both moduli and axions in the low energy theory. 
However, given the theoretical motivation behind axions as both cold dark matter candidates and solving the Strong CP problem of QCD, such couplings may be present in a wide range of theories which also contain an additional massive scalar.

The axions produced through this effect are relativistic at the time of production and thus initially behave as dark radiation.
Due to the high scale of the production, the axions are then redshifted as the universe expands.
Depending crucially on the mass ratio of the modulus to the axion (e.g. the ratio of the scale of production to the momenta at which the axion becomes non-relativistic), we find that the axion may be redshifted to sufficiently low momentum to behave as cold dark matter at the time of BBN.
Additionally, even if the mass ratio is large, the axion contributes only a small amount of dark radiation compared to the background Standard Model radiation resulting from inflationary reheating.
However, if a near-perfect decay efficiency is not achieved through the resonance process, the modulus will eventually dominate the energy density of the universe, and the standard picture of cosmological moduli holds -- with substantial entropy injection effectively erasing all axion production due to the resonance.
From an observational perspective, this makes distinguishing partial non-perturbative decays challenging -- although future study of the effect of such a resonance on axion perturbations may provide a distinction in the production of axion substructures.
Despite this complication, the interplay between moduli decay and axion production offers valuable insight into the cosmological evolution of these fields and the broader implications for dark matter and radiation.

Beginning with the dimension-five operator, we found the instability in the axion field produces an axion growth rate peaking at around $0.5$ for modes with a physical momentum $\tilde{k} \sim m_\phi$.
In addition, when higher-order terms are considered from a UV completion, the growth rate can be enhanced or diminished, and may excite higher momenta modes -- where the result depends on whether the higher-order terms interfere in a constructive or destructive manner.
In particular, models inspired by dilatonic couplings appear to have similar growth rates and instability bands to our heuristic model that contains only the dimension-five term.

We have also numerically simulated the evolution of these fields as the universe expands.
As the initial modulus amplitude approaches the Planck scale, we find that it can lead to a modest increase in the average axion energy density near the modulus oscillation scale. 
In order to trigger rapid non-perturbative moduli decay, however, a large coupling is also required -- which necessitates a more complete UV description as other tachyonic instabilities may arise at low orders which may not be present in the UV.
Additionally, the derived couplings in Large Volume Scenario models are far below the level needed to achieve non-perturbative decay.
As a result, we find that this non-perturbative decay does not drastically change the cosmological picture in Type IIB LVS models.
Furthermore, it appears that sizable demands must be made on string models in order to solve the cosmological moduli problem through non-perturbative decay.

\section*{Acknowledgements}
We thank Mustafa Amin, Amy Burks, Adrienne Erickeck, Nemanja Kaloper and Matt Reece for useful conversations.  S.W. thanks KITP Santa Barbara, the Simons Center and the University of South Carolina for hospitality and research support in part by DOE grant DE-FG02-85ER40237. K.S. thanks the Simons Center and the organizers of the CETUP* and PITT-PACC  workshops for hospitality and is supported in part  by the U.S. National Science Foundation under Grant PHY-2412671.  L.B. and F.C.A. are supported, in part, by the Leinweber Center for Theoretical Physics at the University of Michigan.

\appendix

\section{Semi-analytic treatment of homogeneous modes}
\label{appendix}
In this appendix, we focus on the treatment of the zero modes of the axion and modulus as these play the predominant role in the evolution of the background and the role of the axion as dark matter.
The Lagrangian
\begin{equation}
    \mathcal{L}
    =
    \frac{1}{2}
    \partial_\mu 
    \phi 
    \partial^\mu 
    \phi 
    +
    \frac{1}{2}
    \partial_\mu 
    a
    \partial^\mu 
    a
    +
    \frac{c}{m_P}
    \,
    \phi 
    \partial_\mu
    a
    \partial^\mu 
    a
    -
    \frac{1}{2}
    m_\phi^2
    \phi^2
    -
    \frac{1}{2}
    m_a^2
    a^2
\end{equation}
produces the stress tensor contributions for the axion:
\begin{equation}
    \left(
        T_\mu^\nu
    \right)_a 
    =
    \left( 
        1
        +
        2
        \frac{c}{m_P}
        \phi
    \right)
    \partial_\mu 
    a
    \partial^\nu
    a
    -
    \frac{1}{2}
    \delta_\mu^\nu 
    \left[
        \left( 
            1
            +
            2
            \frac{c}{m_P}
            \phi
        \right)
        \partial_\rho 
        a
        \partial^\rho 
        a
        -
        m_a^2
        a^2
    \right]
\end{equation}
and the modulus:
\begin{equation}
    \left(
        T_\mu^\nu 
    \right)_\phi
    =
    \partial_\mu 
    \phi
    \partial^\nu
    \phi
    -
    \frac{1}{2}
    \delta_\mu^\nu 
    \left[
        \partial_\rho
        \phi 
        \partial^\rho
        \phi 
        +
        2
        \frac{c}{m_P}
        \phi 
        \,
        \partial_\rho
        a
        \partial^\rho
        a
        -
        m_\phi^2
        \phi^2
    \right].
\end{equation}
Focusing on the regime where both the modulus and axion are well-described by their homogeneous components, the axion zero-mode energy density and pressure are respectively given by
\begin{align}
    \rho_{a}
    &=
    \frac{1}{2}
    \dot{a}^2
    +
    \frac{c}{m_P}
    \phi
    \dot{a}^2
    +
    \frac{1}{2}
    m_a^2
    a^2
    \\
    \mathcal{P}_{a}
    &=
    \frac{1}{2}
    \dot{a}^2
    +
    \frac{c}{m_P}
    \phi
    \dot{a}^2
    -
    \frac{1}{2}
    m_a^2
    a^2
\end{align}
so that its equation of state is given by 
\begin{equation}
    w_a
    =
    \frac{
        \dot{a}^2
        +
        2
        \frac{c}{m_P}
        \phi
        \dot{a}^2
        -
        m_a^2
        a^2
    }{
        \dot{a}^2
        +
        2
        \frac{c}{m_P}
        \phi
        \dot{a}^2
        +
        m_a^2
        a^2
    }
    .
\end{equation}
The modulus zero-mode energy density and pressure are given by 
\begin{align}
    \rho_\phi
    &=
    \frac{1}{2}
    \dot{\phi}^2
    -
    \frac{c}{m_P}
    \phi
    \dot{a}^2
    +
    \frac{1}{2}
    m_\phi^2
    \phi^2
    \\
    \mathcal{P}_\phi
    &=
    \frac{1}{2}
    \dot{\phi}^2
    +
    \frac{c}{m_P}
    \phi 
    \dot{a}^2
    -
    \frac{1}{2}
    m_\phi^2
    \phi^2
\end{align}
which produces its equation of state,
\begin{equation}
    w_\phi
    =
    \frac{
        \dot{\phi}^2
        +
        2
        \frac{c}{m_P}
        \phi 
        \dot{a}^2
        -
        m_\phi^2
        \phi^2
    }{
        \dot{\phi}^2
        -
        2
        \frac{c}{m_P}
        \phi
        \dot{a}^2
        +
        m_\phi^2
        \phi^2
    }
    .    
\end{equation}
It is immediately clear that before the onset of axion oscillations (i.e. $H \gtrsim m_a$ so that $\dot{a}\simeq 0$), any modifications to the usual axion and modulus equations of state are negligible.
Once the axion begins to oscillate (i.e. $H\lesssim m_a$), the modulus field $\phi$ has diminished in magnitude by a factor of $H_{a \text{ osc.}}^2 / H_{\phi \text{ osc.}}^2 \sim m_a^2/m_\phi^2 \ll 1$ (assuming modulus domination between both epochs).
Thus even once the axion has begun to oscillate, the modifications to both equations of state are expected to be negligible outside of regimes with significant preheating.

The equations of motion for the homogeneous fields are given by 
\begin{equation}
    \label{eq:eomModulusZeroMode}
    \phi^{\prime \prime}
    +
    \frac{3H}{m_\phi}
    \phi^\prime
    +
    \phi
    =
    \frac{c}{m_P}
    (a^\prime)^2
\end{equation}
and
\begin{equation}
    \label{eq:eomAxionZeroMode}
    a^{\prime \prime}
    +
    \frac{3H}{m_\phi}
    a^\prime
    +
    \frac{m_a^2/m_\phi^2}{1+2\frac{c}{m_P}\phi}
    a
    =
    -
    2
    \frac{c/m_P}{1+2\frac{c}{m_P}\phi}
    \left( 
        \phi^\prime
        a^\prime
    \right)
\end{equation}
where we again use primes to denote differentiation with respect to $z\equiv m_\phi t$.

\subsection{Behavior prior to modulus oscillations}
We first consider the case where the modulus is still frozen by Hubble friction, so that 
\begin{equation}
    \phi \simeq \phi_0.
\end{equation}
In this limit the axion equation of motion reduces to the familiar form of a damped oscillator, producing the solutions
\begin{equation}
    a(z) 
    = 
    k_{1,2}
    \exp
    \left(
        \pm
        z
        \sqrt{\frac{9 H^2}{4m_\phi^2} - \frac{m_a^2/m_\phi^2}{1+2\frac{c}{m_P}\phi_0}} 
        -z
        \frac{3 H}{2m_\phi}
    \right)
\end{equation}
where the integration constants $k_{1,2}$ fix $a(z\sim 0) = a_0$.
While $H\gg m_a$, the axion field is effectively constant, as expected, and the axion begins to oscillate once $m_a \lesssim H$ and if $c \phi_0 / m_P > -1/2$, where $c \phi_0 / m_P$ constitutes only a very small correction to the axion frequency in the absence of the gravitational coupling.
However, for $c \phi_0 / m_P < -1/2$ the effective mass term can be driven tachyonic and produces an exponential growth in the axion field once $H$ is sufficiently low.

\begin{figure}[htb!]
    \centering
    \includegraphics[scale=0.45]{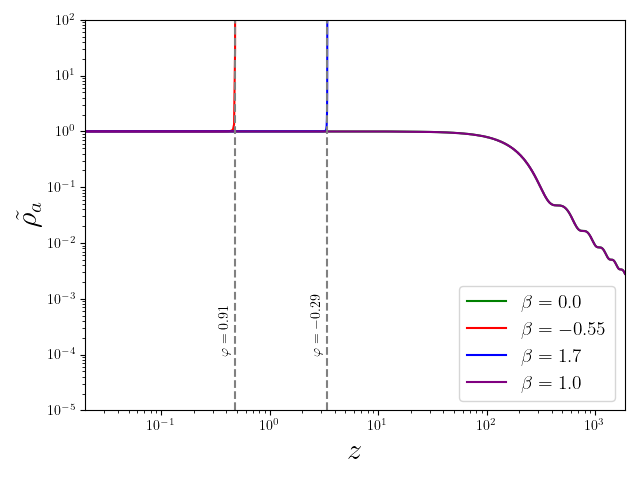}
    \includegraphics[scale=0.45]{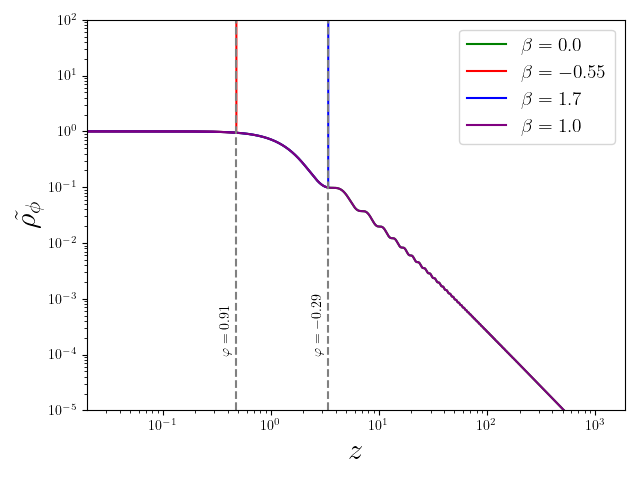}
    \caption{The evolution of the normalized energy densities, $\tilde{\rho}_a = \rho_a / (a_0^2 m_a^2)$ and $\tilde{\rho}_\phi = \rho_\phi / (\phi_0^2 m_\phi^2)$, for the axion (left) and modulus (right) zero-modes.  
    We show the evolution for $\beta\equiv c \phi_0 / m_P = 1$ (purple) which overlaps the case where the gravitational coupling is turned off, $\beta=0$ (green).
    A tachyonic instability is shown for $\beta=-0.55$ (red) and $\beta = 1.7$ (blue). We take $m_\phi/m_a = 100$ to visualize both scales on the same axis.}
    \label{fig:axionRhoUnstable}
\end{figure}

We show this behavior in Figure~\ref{fig:axionRhoUnstable}, which displays the numerical solution to \eqref{eq:eomModulusZeroMode}-\eqref{eq:eomAxionZeroMode}.  
Once the Hubble friction is small and the tachyonic instability has taken over, the exponential increase in the axion field becomes an effective source term in the modulus equation, as can be seen from \eqref{eq:eomModulusZeroMode}, which appears to also drive the modulus energy density towards infinity -- indicating our simplified treatment has entirely broken down.
Close inspection of Figure~\ref{fig:axionRhoUnstable} reveals that the axion instability indeed occurs slightly earlier than that of the modulus, where a full treatment tempers both instabilities through significant production of higher momenta modes.

\subsection{Behavior after onset of modulus oscillations}
To study the behavior after the onset of modulus oscillations, we assume here that the time scale of modulus oscillations is sufficiently small in comparison to the background expansion so that we may parameterize the modulus as before, i.e. taking
\begin{equation}
    \phi = \phi_0 \cos( z )
\end{equation}
where again we define $z\equiv m_\phi t$.
Under this parameterization, we can rewrite \eqref{eq:eomAxionZeroMode} as
\begin{equation}
    \label{eq:axionZeroModeNewTimeVar}
    a^{\prime \prime}
    +
    \left( 
        -
        \frac{
            2
            \frac{c}{m_P}
            \phi_0
            \sin(z)
        }{
            1
            +
            2
            \frac{c}{m_P}
            \phi_0
            \cos(z)
        }
    \right)
    a^{\prime}
    +
    \left(
        \frac{
            m_a^2/m_\phi^2
        }{
            1
            +
            2
            \frac{c}{m_P}
            \phi_0
            \cos(z)
        }
    \right)
    a
    =
    0
\end{equation}
where, again, we use primes to denote differentiation with respect to $z$ and we have taken $3H/m_\phi \simeq 0$.
We now see axion can be driven tachyonic if the condition $|2\frac{c}{m_P} \phi_0| \geq 1$ is satisfied, which may persist once UV corrections have been taken into account.

\begin{figure}[htb!]
    \centering
    \includegraphics[scale=0.7]{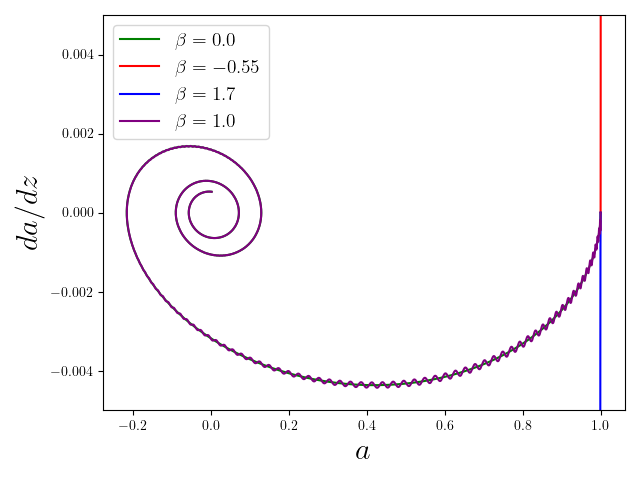}
    \caption{The phase portrait for the axion field over six axion oscillation cycles, in units where $a_0 = 1$. 
    The evolution for $\beta\equiv \frac{c}{m_P}\phi_0 = 1$ (purple) follows a stable trajectory along the path of the $\beta=0$ curve (green), while $\beta = -0.55$ (red) and $\beta = 1.7$ (blue) are unstable.
    We take $m_\phi/m_a = 100$ to visualize both scales on the same axis, and take $a_0 / \phi_0 = 0.001$.}
    \label{fig:axionPhaseSpace}
\end{figure}

The axion equation of motion can also be rewritten as the real part of the following equation:
\begin{equation}
    \label{eq:solveThisOne}
    a^{\prime \prime}
    +
    \left( 
        \frac{
            2
            i
            \frac{c}{m_P}
            \phi_0
            \exp(i z)
        }{
            1
            +
            2
            \frac{c}{m_P}
            \phi_0
            \exp(i z)
        }
    \right)
    a^{\prime}
    +
    \left(
        \frac{
            m_a^2/m_\phi^2
        }{
            1
            +
            2
            \frac{c}{m_P}
            \phi_0
            \exp(i z)
        }
    \right)
    a
    \simeq
    0
\end{equation}
which provides the superimposed analytic solutions for the axion field at late-times:
\begin{equation}
    a(z)
    \simeq
    k_{1,2} 
    \,
    \text{Re}
    \Bigg[
        e^{\pm i z m_a / m_\phi } 
        \,
        _2F_1
        \left(
            1 \pm \frac{m_a}{m_\phi},
            \pm \frac{m_a}{m_\phi};
            1\pm 2\frac{m_a}{m_\phi };
            -2 \frac{c}{m_P} \phi_0 
            e^{i z } 
        \right)
    \Bigg]
    \label{eq:axionFieldSolution}
    .
\end{equation}
Here, $_2F_1(a,b;c;z)$ is the Gaussian hypergeometric function and $k_{1,2}$ are set by initial conditions.
In the phenomenologically expected limit where $m_a/m_\phi \ll 1$, the axion field reduces to the expected simple oscillating term, $a(z) \propto \cos(z m_a/m_\phi) = \cos(m_a t)$, with a much longer time scale than the modulus.
In this limit, we find the axion velocity to be given by
\begin{equation}
    a'(z)
    \simeq
    \pm
    k_{1,2}
    \,
    \text{Re}
    \Bigg[
        i 
        \frac{m_a}{m_\phi}
        e^{\pm i zm_a/m_\phi} 
        \left(
            1
            -
             \frac{
                2
                \frac{c}{m_P}
                \phi_0
                 e^{i z} 
            }{
                1
                +
                2
                \frac{c}{m_P} 
                \phi_0 e^{i z}
            }
        \right)
    \Bigg]
\end{equation}
so that the increase in the axion velocity is immediately apparent if the tachyonic condition $|2\frac{c}{m_P} \phi_0| \geq 1$ is fulfilled.
In Figure~\ref{fig:axionPhaseSpace}, we display the phase portrait for the axion from the numerical solution to \eqref{eq:eomModulusZeroMode}-\eqref{eq:eomAxionZeroMode}, where we have assumed the conventional initial conditions $a^{\prime}(z=0)=0$ and $a(z=0) = a_0 = 1$ in non-dimensional units.

\bibliography{ref}


\end{document}